\documentclass[manuscript]{aastex}
\usepackage{amsmath}
\usepackage{epstopdf}

\shorttitle{Nearby Plasma with Longest Baseline}
\shortauthors{T.V.~Smirnova et al.}

\begin{document}


\title{RadioAstron Studies of the Nearby, Turbulent Interstellar Plasma \\
With the Longest Space-Ground Interferometer Baseline}


\author{T. V. Smirnova\altaffilmark{1},
V. I. Shishov\altaffilmark{1},
M. V. Popov\altaffilmark{2},
C. R. Gwinn\altaffilmark{3},
J. M. Anderson\altaffilmark{4},
A. S. Andrianov\altaffilmark{2},
N.    Bartel\altaffilmark{5},
A.    Deller\altaffilmark{6},
M. D. Johnson\altaffilmark{3},
B. C. Joshi\altaffilmark{7},
N. S. Kardashev\altaffilmark{2},
R.    Karuppusamy\altaffilmark{4},
Y. Y. Kovalev\altaffilmark{2, 4},
M.    Kramer\altaffilmark{4},
V. A. Soglasnov\altaffilmark{2},
J. A. Zensus\altaffilmark{4},
V. I. Zhuravlev\altaffilmark{2}
}
\altaffiltext{1}
{Pushchino Radio Astronomy Observatory, Astro Space Center, Lebedev Physical Institute, 
Russian Academy of Sciences, Pushchino, Moscow oblast', 142290, Russia
}
\altaffiltext{2}
{Astro Space Center, Lebedev Physical Institute, Russian Academy of Sciences,  
Profsoyuznaya str.84/32, Moscow 117997, Russia
\email{zhur@asc.rssi.ru}
}
\altaffiltext{3}
{Department of Physics, University of California, Santa Barbara, California 93106, USA
}
\altaffiltext{4}
{Max-Planck-Institut f\"{u}r Radioastronomie, Auf dem H\"{u}gel 69, 53121 Bonn, Germany
}
\altaffiltext{5}
{York University, Department of Physics and Astronomy, 4700 Keele Street,
Toronto, Ontario M3J 1P3 Canada
}
\altaffiltext{6}
{The Netherlands Institute for Radio Astronomy (ASTRON), Dwingeloo, The Netherlands
}
\altaffiltext{7}
{National Centre for Radio Astrophysics, Post Bag 3, Ganeshkhind, Pune 411007, India
}




\begin{abstract}
RadioAstron space-ground VLBI observations of the pulsar \object{B0950+08}, conducted with the 10-m space 
radio telescope in conjunction with the Arecibo 300-m telescope and Westerbork Synthesis Radio Telescope 
at a frequency of 324 MHz, were analyzed in order to investigate plasma inhomogeneities in 
the direction of this nearby pulsar. The observations were conducted  at a spacecraft distance of 330{,}000 km,
resulting in a projected baseline 
of 220{,}000 km, providing the greatest angular 
resolution ever achieved at meter wavelengths. 
Our analysis is based on fundamental behavior of structure and coherence functions. 
We find that the pulsar shows scintillation on two frequency scales, both much less than the observing frequency;
but modulation is less than 100\%.
We infer that the scattering is weak, but a refracting wedge disperses the scintillation pattern.
The refraction angle of this ``cosmic prism'' is measured as $\theta_0=1.1 - 4.4$ 
mas, with the refraction direction 
being approximately perpendicular to the observer velocity.  
We show that the observed parameters of scintillation effects indicate that two plasma layers lie along the line of sight to the 
pulsar, at distances of $4.4 - 16.4$~pc and $26 - 170$~pc,
and traveling in different directions relative to the line of sight.
Spectra of turbulence  for the two layers are found to follow a power law with the indices  
$\gamma_1 = \gamma_2 = 3.00\pm 0.08$,
significantly different from the index expected for a Kolmogorov spectrum of turbulence, $\gamma=11/3$.
\end{abstract}





\section{Introduction}\label{sec:intro}

Small-scale fluctuations in the electron density of the interstellar 
medium (ISM) scatter radio waves from pulsars. We can study properties of both the ISM and 
the pulsar magnetosphere by measuring frequency-time characteristics of pulsar scintillation 
on the baseline of the cosmic interferometer resulting from scattering. 
Interferometry provides for comparison of scintillation at two different places at the same time.

The goal of the present study is to investigate the spatial distribution 
of scattering along the line of sight to pulsar \object{B0950+08},
one
of the brightest and nearest pulsars. Its distance, measured by
parallax, is $262\pm 5$ pc \citep{br}.
The observations were conducted at a baseline projection of 220{,}000 km and with a spacecraft 
distance of 330{,}000 km providing angular resolution of 1 mas -- the maximum ever achieved 
at meter wavelengths,
on the longest baseline yet attained for VLBI observations,
using the RadioAstron orbiting antenna.

\subsection{Interstellar Scattering Material}

Observations suggest the presence of three components of
scattering material in our galaxy \citep{sm1, sm2}. The first is
unevenly distributed material in the space between spiral arms
(component A). The second corresponds to a cavity of depleted
electron density extending as far as 300 pc from the sun in the
direction perpendicular to the galactic plane and 50 to 100 pc in
the Galactic plane (component B) \citep{snow, bh}. The power
spectra of density fluctuations for components A and B are well
described by a Kolmogorov spectrum with an index of 11/3
\citep{ar, shsm}. The third, component C, is located only about 10 pc from the
sun and has an increased level of turbulence. This component is
responsible for the intra-day variability of quasars at centimeter
wavelengths \citep{dt, ric}. 

As discussed in earlier papers
\citep{sm1, sm2}, the main contributor to the scintillation of the
nearby pulsars \object{J0437-4715} and \object{B0950+08} is
component C. 
The power spectrum of density fluctuations for these
pulsars is flatter than the Kolmogorov spectrum. The index of the
power-law is $\gamma = 3.00 \pm 0.05$ for \object{PSR~B0950+08}
and $\gamma = 3.46 \pm 0.20$ for \object{PSR~J0437-4715}. As shown
by these studies, the level of turbulence in component C is a
factor of 20 higher than that in the extended region responsible for
scintillation of \object{PSR~B0809+74} \citep{sm2}. 
\citet{phi}
assumed that the scattering material in the direction of 
B0950+08 was uniformly distributed, and found that the level of
turbulence an order of magnitude lower than for any previously
measured interstellar line of sight. Their results are also consistent
with enhanced scattering in component C, at a distance of only 10 pc. 

IDV sources show
large-amplitude and rapid variability 
caused by scintillation in component C. The time
scale of these variations is about the same as it would be for
pulsars at centimeter wavelengths.
\citet{Lin} suggested that the partially-ionized surfaces of nearby interstellar clouds
are responsible for intraday
variability.
They note that observed intraday variable sources lie behind such clouds; some where clouds collide.
The nearby scattering screen for \object{PSR~B0950+08} may have the same origin;
indeed, its line of sight passes through one of the clouds they identify.
The far edge of the Local Bubble may also scatter \object{PSR~B0950+08}.
In this direction, the edge of the Local Bubble lies at a distance of 120 to 160 pc  \citep{lal}.
\citet{sm2}
suggested the existence of strong angular refraction in the
direction to \object{PSR~B0950+08}.

\subsection{Orbiting Antenna and Observations}

Interferometric observation of pulsars with space and Earth antennas, 
for baselines of several Earth diameters,
provides the possibility of
localizing the scattering layers and also allows one to evaluate the influence of refraction on 
the received emission. \citet{sh2} presents a preliminary theoretical analysis of this approach.
For the first such VLBI observations of the Crab pulsar with ground telescopes see \citet{kon}.
As we discuss in Section\ \ref{sec:FresnelScales},
the typical scale of the scattering pattern of a weakly-scattered pulsar is the Fresnel scale,
larger than an Earth diameter.
Space-Earth interferometry affords the possibility of measuring this scale directly,
rather than allowing motions of pulsar, Earth and scattering material to carry it across a single antenna.
We make use of both interferometric and time-lag analyses in this paper;
the earlier work by \cite{sm2} used only single-dish observations and provides an interesting comparison.

Because scintillation is a stochastic process, observations must be compared with 
theoretically-predicted distributions.
These distributions may not have analytical forms and can be quite difficult to compute \citep[e.g.,][]{gw,jo}.
Consequently, moments of the distributions, particularly in the form of structure functions,
provide measures of the fundamental behavior of structure and coherence functions.
Section \ref{sec:ISMScatteringTheory} describes this approach in the present case.

In this paper, we present results obtained from observations of the pulsar \object{PSR~B0950+08} 
carried out on 25 January 2012 at a frequency of 324 MHz with the RadioAstron space radio telescope in 
conjunction with the Arecibo 300-m telescope (AR)
and the Westerbork synthesis array (WB).
The RadioAstron project is an international collaborative mission involving a free-flying 
satellite, Spectr-R, carrying a 10-m space radio telescope (SRT) 
on an elliptical orbit around the Earth. This space telescope 
performs radio astronomical 
observations using very long baseline interferometry (VLBI) techniques in conjunction with 
ground-based VLBI networks. The orbit of the RadioAstron satellite evolves with time. It has 
an apogee between 280,000 and 350,000 km, a perigee between 7,000 and 80,000 km, a period of 
8 to 9 days, and an initial inclination of $51^{\circ}$. RadioAstron operates at the standard radio 
astronomical wavelengths of 1.19 -- 1.63 cm (K-band), 6.2 cm (C-band), 18 cm (L-band), and 92 cm 
(P-band). Technical parameters of the on-board scientific equipment and measured parameters have been 
described in two main publications \citep{ak, kar}.

\section{ISM Scattering Theory}\label{sec:ISMScatteringTheory}
\label{main}
\subsection{Overview of Model}

As we discuss in Section\ \ref{sec:analysis_and_results}, 
pulsar \object{B0950+08} shows scintillation with modulation $\Delta I/I = m <1$,
with typical bandwidth $\Delta\nu$ much less than the observing frequency $\nu$.
Typically $m<100\%$ is observed only in weak scattering,
whereas $\Delta\nu<\nu$ is observed only in strong scattering \citep{cc}.
These two regimes are distinct.
In weak scattering the difference in phase among paths from source to observer is less than $\pi$ radians,
but in strong scattering it is more.

We suggest that a strong gradient in the column density of refracting material -- a prism -- is responsible for the apparent paradox.
Such a prism will disperse the wide-band scintillation pattern in the observer plane,
so that maxima and minima appear at different places at different observing frequencies.
Indications of strong refractive effects include the typically ``refractive'' scaling of scintillation bandwidth with frequency found by \citet{sm2},
and the shift of the structure function of visibility with frequency, as a function of time lag, discussed in Section\ \ref{sec:asymmetry} and 
displayed in Figure\ \ref{fig.8}.
\citet[][see their Figure 10]{sm2} showed that the transition between weak and strong scattering for this pulsar takes place in the frequency range of 100 to 300~MHz,
with of course strong scattering at lower frequencies.

We find that 2 screens at different distances are required to reproduce the observed properties of scintillation.
In particular, we find that the structure function of visibility with observing frequency is composed of two components with different timescales
and different behavior as a function of interferometer baseline.
We discuss the structure function in Section\ \ref{sub1.1},
and present the structure functions formed from our observations in Section\ \ref{sec:estimated_structure_function}.
Because the scintillation is weak, the observed scintillation pattern in the observer plane is the superposition of the patterns from the two screens.




\subsection{Characteristics of the ISM structure function}
\label{sub1.1}

This paper considers interferometric observations of scattering.
The interferometric visiblity $V$ is as a function of frequency, time and baseline is the fundamental observable.
We are concerned with the fluctuations of $V$.
We analyze these fluctuations using the fundamental behavior of structure and coherence functions.
As we describe in this section, we relate the modulus of the visibility, the product of electric fields at two positions,
to the product of the intensities at those positions.
The resulting expressions are excellent when the noise level is low. In the case of a high 
level of noise, it is advantageous to measure fluctuations in the \emph{squared} modulus of the interferometer cross-power spectrum $V^2$. 
In Appendix A, we derive the analogous relations that account for 
noise. 
In particular, we show that for statistics of $V^2$ the contributions of noise and signal are simply additive,
whereas
for statistics of $V$ or $|V|$ the contributions of noise and signal are much more difficult to separate.
We therefore deal with the statistics of $V^2$ in this paper.

\subsubsection{Field and propagation}
We define $h(f,t)$ as the electric field of pulsar emission, in the absence of a turbulent 
plasma, where $f = \nu - \nu_0$ is the offset of the observing frequency $\nu$ from the band center $\nu_0$, and $t$ is time.  
This electric field $h(f,t)$ also includes modulation by the receiver bandpass. The field after 
propagation through the ISM can be represented as \citep{sh2}
\begin{eqnarray}
E({\boldsymbol\rho},f,t)=u({\boldsymbol\rho},f,t)h(f,t),
\label{eq:edef}
\end{eqnarray}
where scattering in the ISM results in the factor $u({\boldsymbol\rho},f,t)$, given that ${\boldsymbol\rho}$ is the spatial 
coordinate in the observer plane perpendicular to the line of sight. 
Here and throughout the paper,
we use boldface type to denote vector quantities, such as ${\boldsymbol\rho}$.
To obtain the cross-power spectrum $V$ -- i.e., the 
response of an interferometer with baseline $\Delta{\boldsymbol\rho}$ averaged over a fixed realization of the scattering -- 
we multiply $E({\boldsymbol\rho},f,t)$ by 
$E^{\ast}({\boldsymbol\rho} + \Delta{\boldsymbol\rho},f,t)$ and average over statistics of the source electric field:
\begin{eqnarray}
V({\boldsymbol\rho},{\boldsymbol\rho}+\Delta{\boldsymbol\rho},f,t)&=&\Big\langle E({\boldsymbol\rho},f,t)E^{\ast}({\boldsymbol\rho}+
\Delta{\boldsymbol\rho},f,t)\Big\rangle_h \nonumber\\
&=& j({\boldsymbol\rho},{\boldsymbol\rho}+\Delta{\boldsymbol\rho},f,t)H(f,t),
\label{eq:vdef}
\end{eqnarray}
where
\begin{eqnarray}
j({\boldsymbol\rho},{\boldsymbol\rho}+\Delta{\boldsymbol\rho})&=&u({\boldsymbol\rho},f,t)u^{\ast}({\boldsymbol\rho}+\Delta{\boldsymbol\rho},f,t)\\
\nonumber H &=&\Big\langle h(f,t)h^{\ast}(f,t)\Big\rangle_h .
\label{eq:jdef}
\end{eqnarray}
Here, the angular brackets $\langle \ldots \rangle_h$ with subscript $h$ indicate an average over the noiselike statistics of the electric field of the source. 
The flux density of the source, corrected for bandpass, is $H$; and the effects of scattering are expressed by $j$.
Note that $H$ depends on frequency within the observed band $f$ and on time $t$, but we omit these arguments
for clarity in the equations below.

\subsubsection{Visibility and fluctuations}\label{sec:visbilityandfluctuations}

Consider fluctuations of the modulus of the interferometer response (the dynamic cross-power spectrum):
\begin{eqnarray}
\left| V({\boldsymbol\rho},{\boldsymbol\rho}+\Delta{\boldsymbol\rho},f,t)\right|&=&
H\cdot {\left[I({\boldsymbol\rho})I^{\ast}({\boldsymbol\rho}+\Delta{\boldsymbol\rho})\right]}^{1/2},
\label{eq:relate_V_and_I}
\end{eqnarray}
where the intensity $I$ is the square modulus of electric field at a single position:
\begin{eqnarray}
I({\boldsymbol\rho})=j({\boldsymbol\rho},{\boldsymbol\rho})=u({\boldsymbol\rho},f,t)u^{\ast}({\boldsymbol\rho},f,t) . \label{eq:Idef}
\end{eqnarray}
Again, note that $H$, $j$, and $I$ depend on $f$ and $t$, but we omit these arguments for simplicity.
We normalize the average flux density of the source, so that the average over realizations of scintillations is unity:
$\langle I({\boldsymbol\rho}) \rangle_{s}=1$. We are concerned with the fluctuations of intensity about this average: $\Delta I = I - 1$.

In the regime of weak scintillation, $|\Delta I({\boldsymbol\rho})|\ll 1$, so $| V({\boldsymbol\rho},{\boldsymbol\rho}+\Delta{\boldsymbol\rho})|$ can be approximated as
\begin{eqnarray}
&&\left| V({\boldsymbol\rho},{\boldsymbol\rho}+\Delta{\boldsymbol\rho},f,t)\right| \\
&&\quad\quad \approx
H\cdot \left( 1+\textstyle{\frac{1}{2}}\Delta I({\boldsymbol\rho})+\textstyle{\frac{1}{2}}\Delta I({\boldsymbol\rho}+\Delta{\boldsymbol\rho})\right).
\nonumber
\end{eqnarray}
Thus, fluctuations in the interferometric visibility $| V({\boldsymbol\rho},{\boldsymbol\rho}+\Delta{\boldsymbol\rho},f,t)|$ are given by
\begin{eqnarray}
&&\left| \Delta V({\boldsymbol\rho},{\boldsymbol\rho}+\Delta{\boldsymbol\rho},f,t) \right| \\
&&\quad\quad =
\left| V({\boldsymbol\rho},{\boldsymbol\rho}+\Delta{\boldsymbol\rho},f,t)\right| -\big\langle \left| V({\boldsymbol\rho},{\boldsymbol\rho}+\Delta{\boldsymbol\rho},f,t)\right|\big\rangle_s \nonumber\\
&&\quad\quad\approx 
H \cdot \left( \textstyle{\frac{1}{2}} \Delta I({\boldsymbol\rho})+\textstyle{\frac{1}{2}} \Delta I({\boldsymbol\rho}+\Delta{\boldsymbol\rho})\right).
\nonumber
\end{eqnarray}

\subsubsection{Screens and statistics}\label{sec:screensandstatistics}

We suppose that the scattering material lies in two phase-changing screens located at distances $z_1$ and $z_2$ from the observer. 
The distance of the source form the observer is $z$.
Each screen produces some variation of phase as a function of position.
We can characterize the statistics of the phase screen
by two spatial structure functions of phase fluctuations $D_{S,1}(\Delta {\mathbf x}_1)$ and $D_{S,2}(\Delta {\mathbf x}_2)$,
where $\Delta {\mathbf x}_1$ and $\Delta {\mathbf x}_2$ are the differences of the spatial coordinates in the plane of the phase-changing screen:
\begin{eqnarray}
D_{S,\ell}(\Delta {\mathbf x}_\ell) &=& \Big\langle \big( \Phi_\ell( {\mathbf x}_\ell) -  \Phi_\ell( {\mathbf x}_\ell- \Delta {\mathbf x}_\ell ) \big)^2 \Big\rangle_s
\label{eq:Dsdef}
\end{eqnarray}
where $\ell=1,2$ identifies the screeen, and $\Phi_\ell({\mathbf x}_\ell)$ is the screen phase at ${\mathbf x}_\ell$.
Here, the angular brackets with subscript $s$ indicate an average over an ensemble of statistically-identical scattering media.

\citet{sm2} showed that the spectrum of turbulence of the interstellar medium in the direction of 
\object{PSR~B0950+08} has a power-law form,
$\Phi_{S} (q)\propto |q|^{\alpha+2}$, where $q$ is spatial frequency.
Consequently, the spatial structure
functions of phase fluctuations exhibit a power-law form as well. 
We describe the
structure functions of the two phase-changing screens with the expressions
\begin{eqnarray}
D_{S,\ell}(\Delta {\mathbf x}_\ell)=(k\Theta_{scat,\ell} | {\Delta x_\ell } | )^{\alpha_\ell}  \label{eq:phase_structure_fns}
\end{eqnarray}
where $k = 2\pi/\lambda$ is the wavenumber, $\lambda=c/\nu_0$ is the wavelength, and $\Theta_{scat,\ell}$ represents the angle of scattering at 
phase-changing screen $\ell=1,2$.
We assume a power-law form for the structure functions, with indices $\alpha_1$, $\alpha_2$.   
Note that this equation introduces the assumption that scattering is isotropic.
We make this assumption for the rest of the paper.
In principle, effects of anisotropy could be detected by comparing results on several baselines of comparable lengths and different orientations.
Thus, ongoing and future observations should be able to refine the results presented here.

As was pointed out in \citet[see figure~\ref{fig.10}]{sm2}, the transition
from strong to weak scintillation takes place in the frequency range 100 - 300~MHz.
We will show in section~\ref{sub4.1} that modulation index $m$ at frequency 324~MHz is 
about 0.35, so the scintillation is weak \cite{mar}. In other words, the fluctuations in phase introduced from propagation
are small: $\Delta\Phi_\ell \ll 2 \pi$.
The modulation indices resulting from scattering at the screens are $m_1$ and $m_2$, defined 
formally in Equations\ \ref{q21} and\ \ref{q24} below.  
We can apply weak scattering theory because for a power-law
spectrum of turbulence, the difference between approximate and exact values 
of $m^2$ doesn't exeed of $0.2 m^2$ \citep{mar}.

For a nearby screen at distance $z_1\ll z$, we can use the plane wave approximation, so that
\begin{eqnarray}
D_{S,1}(\Delta {\boldsymbol\rho}) &=& D_{S}(\Delta {\mathbf  x}_1) .
\end{eqnarray}
For a phase-changing screen at distance $z_2$,
the structure function of phase fluctuations at the screen is related to that at the observer by:
\begin{eqnarray}
D_{S,2}(\Delta {\boldsymbol\rho}(z-z_2)/z) &=& D_{S}(\Delta{\mathbf x}_2) .
\end{eqnarray}
Note that these equations describe the effects of the screen as a ``shadowgraph'', where effects of the screen are projected directly onto the observer plane.
This is characteristic of weak scattering.

\subsubsection{Cosmic prism}\label{sec:cosmicprism}

We suppose that a cosmic 
prism, or gradient of refracting material, is located between the pulsar and the phase-changing screens. 
This prism deflects radiation from the pulsar at a frequency-dependent refractive angle.
We parametrize this refraction using ${\boldsymbol\theta}_0$, the apparent displacement of the source location as observed from the observer plane
at frequency $\nu_0$. 
Similar strong angular refraction has been detected in the direction to \object{PSR~B0329+54} 
from analysis of multi-frequency observations \citep{sss73}. 
\citet{sm2} showed that strong angular refraction exists in the direction to
\object{PSR~B0950+08}, so we assume
that the refractive angle is significantly greater than the scattering 
angle at either screen:
\begin{eqnarray}
| {\boldsymbol\theta}_0 | &\gg& \Theta_{scat,1}\\
| {\boldsymbol\theta}_0 | &\gg& \Theta_{scat,2}(z-z_2)/z \nonumber
\end{eqnarray}
Thus, our model of turbulent interstellar plasma in the direction of the pulsar is characterized by 
the following parameters: ${\boldsymbol\theta}_0$, $\alpha_1$, $\alpha_2$, $m_1$, $m_2$, $z_1$, $z_2$.

\subsubsection{Shifts in frequency, position, and time}\label{sec:shifts_in_freq_and_time}

The cosmic prism dominates the angular deflection of the source,
and in particular is greater than that produced by the screens.
In the presence of angular refraction, with the dispersion produced by the interstellar plasma,
a change in frequency of $f$ from the fiducial frequency $\nu_0$
leads to an apparent displacement of the source position by an angle
\begin{eqnarray}
{\boldsymbol\theta}_f= \left( 1 - \frac{\nu_0^2}{(f+\nu_0)^2} \right) \theta_0 \approx 2(f/\nu_0){\boldsymbol\theta}_0 , \label{q15}
\end{eqnarray}
where $\nu_0$ is the frequency at the center of the observing band.

The combination of a cosmic prism and a phase-changing screen leads to a shift in the scintillation pattern.
For a phase-changing screen at distance $z_1$ from the observer, and the cosmic prism beyond the screen,
the apparent displacement of the source leads to a displacement of the scintillation pattern in the observer plane by a distance ${\boldsymbol\rho}_{f,1}$, 
given by \citep{lit, sh1}
\begin{eqnarray}
{\boldsymbol\rho}_{f,1}=z_1{\boldsymbol\theta}_f . \label{q16}
\end{eqnarray}
The further away the screen, the greater the 
dispersion of the scintillation pattern at the Earth produced by the prism beyond the screen.

On the other hand, if the observer travels at velocity ${\mathbf V}_{obs}$ perpendicular to the line of sight, 
and if screen 1 moves at speed ${\mathbf V}_{scr,1}$,
then observer's spatial displacement relative to the scintillation pattern increases with the change of time $\Delta t$ at velocity ${\mathbf V}_{1}$:
\begin{eqnarray}
{\boldsymbol\rho}_{t,1}&=& {\mathbf V}_{1} \Delta t \label{eq:obs_speed}\\
&=& \left({\mathbf V}_{obs}-{\mathbf V}_{scr,1}\right) \Delta t.
\label{eq:time_shift_1}
\end{eqnarray}
If this displacement is parallel to the dispersion of the cosmic prism,
then the observer will notice a shift of the scintillation pattern in frequency, as a function of time.

For a more distant screen, at distance $z_2$ from the observer, the spherical form of the waves must be taken into account:
\begin{eqnarray}
{\boldsymbol\rho}_{f,2}=\frac{z\; z_2}{(z-z_2)}{\boldsymbol\theta}_f,  
\label{eq:f_shift_2}
\end{eqnarray}
and
\begin{eqnarray}
{\boldsymbol\rho}_{t,2}&=& {\mathbf V}_{2} \Delta t \\
&=& \left( {\mathbf V}_{obs} -\frac{z}{(z-z_2)}{\mathbf V}_{scr,2}+\frac{z_2}{(z-z_2)}{\mathbf V}_{PSR} \right) \Delta t. \label{eq:time_shift_2}
\end{eqnarray}
where ${\mathbf V}_{scr,2}$ is the velocity of screen 2, and ${\mathbf V}_{PSR}$ is the velocity of the pulsar.

Thus, a change in frequency will cause a change in the diffraction pattern analogous to a change in position in the direction of ${\boldsymbol\theta}_0$;
and a delay in time is equivalent to a change in position in the direction of a linear combination of ${\mathbf V}_{obs}$, ${\mathbf V}_{scr}$, and ${\mathbf V}_{PSR}$, for a moving observer, screen, or pulsar.
These equivalences arise because the intensity variations from weak scintillation have wide intrinsic bandwidth:
they are dispersed only by the cosmic prism.
We apply these equivalences further in Section\ \ref{sec:time_freq_baseline} below.

\subsubsection{Formation of structure functions}\label{sec:theory_structure_functions}

We construct the structure function of intensity in the observer plane:
\begin{eqnarray}
D_{\Delta I}( \Delta {\boldsymbol\rho} ) &=& \Big\langle \big( \Delta I ({\boldsymbol\rho}+\Delta{\boldsymbol\rho} ) -  \Delta I({\boldsymbol\rho} ) \big)^2 \Big\rangle_s \label{eq:defineIstructurefunc} \\
&=& D_{\Delta I,1}( \Delta{\boldsymbol\rho} ) + D_{\Delta I,2}( \Delta{\boldsymbol\rho} ) \label{eq:screenIstructurefunc}
\end{eqnarray}
In weak scattering, the structure functions of the two screens add to produce the observed structure function,
as we indicate here.
We wish to relate this to the structure function for interferometric visibility.

Consider observations of intensity at two locations in the observer plane separated by $\Delta{\boldsymbol\rho}$,
and separated in frequency by $\Delta f$ and in time by $\Delta t$.
The structure function characterizing the fluctuations of intensities is:
\begin{equation}
\begin{array}{l}
D_{\Delta I,\ell}(\Delta{\boldsymbol\rho},\Delta f,\Delta t)   \\
=
\Big\langle \Big( \left| I_\ell({\boldsymbol\rho},{\boldsymbol\rho}+\Delta{\boldsymbol\rho},f+\Delta f,t+\Delta t)\right|-\left| I_\ell({\boldsymbol\rho},{\boldsymbol\rho},f,t) \right| \Big)^2\Big\rangle_s\\
=H^2 \Big[ D_{\Delta| j|,\ell}({\boldsymbol\rho}_{f,\ell}+{\boldsymbol\rho}_{t,\ell})+
\textstyle{\frac{1}{2}} D_{\Delta| j|,\ell}(\Delta{\boldsymbol\rho}+{\boldsymbol\rho}_{f,\ell}+{\boldsymbol\rho}_{t,\ell}) \\
\phantom{=H^2 \Big[}
+\textstyle{\frac{1}{2}} D_{\Delta| j|,\ell}(\Delta{\boldsymbol\rho}-{\boldsymbol\rho}_{f,\ell}-{\boldsymbol\rho}_{t,\ell})-D_{\Delta| j|,\ell}(\Delta{\boldsymbol\rho})\Big]  . 
\end{array}
\label{q18} 
\end{equation}
where $\ell=1$ or 2 identifies the screen.  
This equation relates the structure functions of intensity to those of interferometric visiblity.
A similar equation provides the reverse relation:
\begin{equation}
\begin{array}{l}
H^2 D_{\Delta| j|,\ell}(\Delta{\boldsymbol\rho},\Delta f,\Delta t) \\
\quad = 
D_{\Delta I,\ell}({\boldsymbol\rho}_{f,\ell}+{\boldsymbol\rho}_{t,\ell})+
\textstyle{\frac{1}{2}} D_{\Delta I,\ell}(\Delta{\boldsymbol\rho}+{\boldsymbol\rho}_{f,\ell}+{\boldsymbol\rho}_{t,\ell})\\
\quad\quad +\textstyle{\frac{1}{2}} D_{\Delta I,\ell}(\Delta{\boldsymbol\rho}-{\boldsymbol\rho}_{f,\ell}-{\boldsymbol\rho}_{t,\ell})
- D_{\Delta I,\ell}(\Delta{\boldsymbol\rho})
\end{array}
\end{equation}
Also note that in the short-baseline limit,
the structure function for the interferometric visibility is that of intensity, with the correction for bandpass.

\subsubsection{Equivalence of time, frequency and baseline}\label{sec:time_freq_baseline}

We can use the results of the previous section to infer dependence of the structure function on time and frequency
as well as baseline,
by noting that the cosmic prism renders a change in time or frequency {\it equivalent} to a change in position.
We can therefore define a generalized position variable ${\mathbf r}_{\ell}$,
which includes effects of dispersion on time and frequency behavior of the scintillation pattern for each screen:
\begin{eqnarray}
{\mathbf r}_{\ell}&=& {\boldsymbol\rho}_{f,\ell} + {\boldsymbol\rho_{t,\ell}} + \Delta{\boldsymbol\rho} \label{eq:r_def}
\end{eqnarray}
Because we assume that the scattering in the screens is isotropic,
the structure function for each screen depends only on the magnitude of its argument:
\begin{eqnarray}
D_{\Delta I,\ell}({\mathbf r}_\ell) &=& D_{\Delta I,\ell}( | {\mathbf r}_\ell |) 
\end{eqnarray}
The direction of refractive dispersion divides the components of the other vectors into those parallel and perpendicular
to the direction of that dispersion. 
The components of the vectors may be different because the conversion of $f$ and $\Delta t$ to ${\boldsymbol\rho}_{f,\ell}$ and ${\boldsymbol\rho}_{t,\ell}$ differs for the screens.
\begin{eqnarray}
\left| {\mathbf r}_{\ell} \right|&=& \Big(  \big| {\boldsymbol\rho}_{f,\ell} \big|^2 + \big| {\boldsymbol\rho_{t,\ell}} \big|^2  +  \big| \Delta{\boldsymbol\rho} \big|^2
\quad + 2  \left| {\boldsymbol\rho}_{f,\ell} \right| \left| {\boldsymbol\rho_{t,\ell}} \right| \cos\beta_\ell    
\label{eq:beta_phi_def} \\
&&
+ 2 \left| {\boldsymbol\rho}_{f,\ell} \right| \left| \Delta{\boldsymbol\rho} \right| \cos\varphi
\quad + 2 \left| {\boldsymbol\rho}_{t,\ell} \right| \left| \Delta{\boldsymbol\rho}  \right| \cos(\beta_\ell - \varphi)  
\Big)^{1/2} \nonumber
\end{eqnarray}
Here, $\beta_\ell$ is the angle between 
the direction of dispersion ${\boldsymbol \theta}_0$, and the velocity ${\mathbf V}_\ell$ of the ray relative to screen $\ell$;
and $\varphi$ is the angle between ${\boldsymbol\theta}_0$ and the baseline $\Delta{\boldsymbol\rho}$.

Physically, this equation states that the structure function for variations in intensity is that set by optics of a static screen without refraction $D_{\Delta| j|,\ell}(\Delta{\boldsymbol\rho})$;
however, refraction and motion introduce correlation in frequency set by the scale ${\boldsymbol\rho}_{f,\ell}$ and in time set by the scale ${\boldsymbol\rho}_{t,\ell}$.
Note that the effects of delays in time or changes in frequency depend sensitively on the angles $\beta_{\ell}$.
Note slso that $D_{\Delta I,\ell}(\Delta \rho, \Delta f, \Delta t)$ is a symmetrical function of $\cos(\varphi)$,
as Equation\ (\ref{q18}) shows. Therefore $D_{\Delta I,\ell}(\Delta\rho, \Delta f, \Delta t = 0)$ is a symmetrical function of $\Delta f$ and $D_{\Delta I,\ell}(\Delta \rho, \Delta f = 0, \Delta t)$ is a symmetrical function of $\Delta t$.\\[40pt]


\subsubsection{Fresnel scales}\label{sec:FresnelScales}

The Fresnel scale for a nearby screen at distance $z_1$ is given by
\begin{eqnarray}
\rho_{Fr,1} = (z_1/k)^{1/2},
\label{eq:Fresnel_close}
\end{eqnarray}
For a nearby screen,
the structure function of intensity variations is given by the simple relations \citep{pr}:
\begin{equation}
\begin{array}{rcll}
D_{\Delta| j|,1}(\Delta{\boldsymbol\rho})&=&2 D_{S,1}(\Delta{\boldsymbol\rho}) & {\rm for\ }\Delta\rho<\rho_{Fr,1} \\
D_{\Delta| j|,1}(\Delta{\boldsymbol\rho})&=& 2 D_{S,1}(\rho_{Fr,1})\approx 2 m^2_1 & {\rm for\ }\Delta\rho>\rho_{Fr,1}
\end{array}
\label{q21}
\end{equation}
where $m_1$ is the modulation index of the nearby layer. 

In contrast, for a distant phase-changing screen, we must account for sphericity of the wave. 
Then, the Fresnel scale of the screen $\rho_{Fr,2}$ 
is given by \citet{lit, sh1} as:
\begin{eqnarray}
\rho_{Fr,2}=[z\; z_2/(z-z_2)k]^{1/2}. 
\label{eq:Fresnel_far}
\end{eqnarray}
Consequently, for a distant scattering screen,
\begin{equation}
\begin{array}{rcll}
D_{\Delta| j|,2}(\Delta{\boldsymbol\rho})&=&2 D_{S,2}\left([(z-z_2)/z]\Delta{\boldsymbol\rho}\right)  
&\rm{for} \quad \Delta\rho<\rho_{Fr,2} \\
D_{\Delta| j|,2}(\Delta{\boldsymbol\rho})&=& 2 D_{S,2}(\rho_{Fr,2})\approx 2 m^2_2
&\rm{for} \quad \Delta\rho>\rho_{Fr,2}.
\end{array}
\label{q24}
\end{equation}
where $m_2$ is the modulation index of a distant scattering screen.
Physically, Equations\ (\ref{q21}) and (\ref{q24}) represent the fact that the diffraction pattern from a static, nonrefracting screen in weak scattering 
has a scale equal to the Fresnel scale.
Together with Equation\ (\ref{q18}), these equations describe the statistics of the diffraction pattern from a screen in the presence of a cosmic prism,
and with motions of source, screen, or observer.

\subsubsection{Fresnel scales for frequency and 
time}\label{sec:Fresnel_scales_freq_time}

The Fresnel scales given by Equations\ (\ref{eq:Fresnel_close}) and\ (\ref{eq:Fresnel_far})
give rise to corresponding scales in frequency 
and time 
through the action of the cosmic prism,
which relates frequency shifts and time lags with changes in position via Equations\ (\ref{q16}) through (\ref{eq:time_shift_2}).
Because the expressions are important in comparing theory with observation, we present expressions for these scales here.
The Fresnel frequency scales are:
\begin{eqnarray}
f_{Fr,1} &=&\frac{\nu_0}{2 \theta_0} \frac{\rho_{Fr,1}}{z_1} =  \frac{\nu_0}{2\theta_0} \sqrt{\frac{1}{k\, z_1}} \label{eq:fFr1_def}\\
f_{Fr,2} &=&\frac{\nu_0}{2 \theta_0} \frac{\rho_{Fr,2}(z-z_2)}{z z_2} = \frac{\nu_0}{2\theta_0} \sqrt{\frac{z-z_2}{k\,  z\, z_2}} \label{eq:fFr2_def}
\end{eqnarray}
for a nearby and more distant screen, respectively.
Note that the frequency scale is largest for a nearby screen, and decreases with increasing screen distance.
The Fresnel time scales are simply:
\begin{eqnarray}
t_{Fr,1} &=& \frac{ \sqrt{z_1/k} }{\left| {\mathbf V}_{obs} \right|} \label{eq:tFr1_def}\\
t_{Fr,2} &=& \frac{ \sqrt{z\, z_2/(k(z-z_2))}}{\left| {\mathbf V}_{obs}+{\mathbf V}_{PSR}(z_2/(z - z_2))\right| } \label{eq:tFr2_def}
\end{eqnarray}
Here, we assume that the velocities of the screens are small: $V_{scr,1},V_{scr,1}\ll V_{obs}$ and $V_{PSR}$.

\subsection{Temporal Coherence Function}\label{sec:tempcoherencefunction}

The inverse Fourier transform of $I({\boldsymbol\rho},{\boldsymbol\rho}+\Delta{\boldsymbol\rho},f,t)$ gives the temporal coherence function, averaged over 
statistics of the source electric field:
\begin{eqnarray}
P({\boldsymbol\rho},{\boldsymbol\rho}+\Delta{\boldsymbol\rho},\tau,t)=\int d f \exp(2\pi i f \tau)I({\boldsymbol\rho},{\boldsymbol\rho}+\Delta{\boldsymbol\rho},f,t).
\label{eq:tempcoherfuncdef}
\end{eqnarray}
where $\tau$ is a time lag of coherence function.
Then, the value of $P$ averaged over the statistics of the turbulent medium is 
\begin{eqnarray}
\langle P({\boldsymbol\rho},{\boldsymbol\rho}+\Delta{\boldsymbol\rho},\tau,t)\rangle_s =B_u(\Delta{\boldsymbol\rho})P_H(\tau,t),
\end{eqnarray}
where 
\begin{eqnarray}
P_H(\tau,t)=\int d f \exp(2 \pi i f \tau)H(f,t),
\label{eq:PHdef}
\end{eqnarray}
and 
\begin{eqnarray}
B_u(\Delta{\boldsymbol\rho})=
\exp\left\{-\frac{1}{2}D_{S,1}[\Delta{\boldsymbol\rho}]-\frac{1}{2}D_{S,2}[\frac{(z-z_2)}{z}]\Delta{\boldsymbol\rho}\right\} \label{q29} 
\end{eqnarray}
Here, $P_H(\tau,t)$ is the temporal coherence function defined by the source and $B_u(\Delta{\boldsymbol\rho})$ is the 
spatial coherence function of the scattered field.

Given these functions, we can characterize fluctuations of $P({\boldsymbol\rho},{\boldsymbol\rho}+\Delta{\boldsymbol\rho},\tau,t)$ by its second moment:
\begin{eqnarray}
&& \langle | P({\boldsymbol\rho},{\boldsymbol\rho}+\Delta{\boldsymbol\rho},\tau,t)|^2 \rangle_s=\nonumber\\
&& \int d f\int d \Delta f\exp(-2 \pi i \Delta f \tau)H(f,t)H(f+\Delta f,t)\nonumber\\
&&\quad \times \langle j({\boldsymbol\rho},{\boldsymbol\rho}+\Delta{\boldsymbol\rho},f,t)j^{\ast}({\boldsymbol\rho},{\boldsymbol\rho}+\Delta{\boldsymbol\rho},f+\Delta f,t)\rangle_s.
\end{eqnarray}
Then, we can write the mean squared modulus of $P$, averaged over statistics of turbulent medium, as
\begin{eqnarray}
\langle | P(\tau)|^2 \rangle_s=\langle | P_0(\tau)|^2 \rangle_s + \langle | P_S(\tau)|^2\rangle_s, \label{q31}
\end{eqnarray}
where $\langle | P_0(\tau)|^2 \rangle_s$ corresponds to unscattered emission
$\langle | P_0(\tau)|^2\rangle_s=\langle P({\boldsymbol\rho},{\boldsymbol\rho}+\Delta{\boldsymbol\rho},\tau,t)\rangle_s^2$, 
and  $\langle | P_S(\tau)|^2\rangle_s$ consists of two parts that give the contributions 
of the nearby and distant screens:
\begin{eqnarray}
\langle | P_S(\tau)|^2 \rangle_s = \langle | P_{S,1}(\tau) |^2\rangle_s + \langle | P_{S,2}(\tau)|^2\rangle_s.
\end{eqnarray}
Component $\langle | P_{S,1}(\tau)|^2\rangle_s$ can be written as \citep{pr, sh1}
\begin{eqnarray}
\langle | P_{S,1}(\tau) |^2 \rangle_s=
\frac{H^2_0\pi\nu_0}{z_1\theta_0}\int d 
q_{\perp}\Phi_{S,1}(q_{||}=\pi\nu_0\tau/z_1\theta_0,q_{\perp}),
\label{eq:PS1def}
\end{eqnarray}
where $H_0$ is the flux density of the source integrated over frequency, $H_0 =\int d f H(f)$, $\Phi_S$ is a power
phase spectrum, and $q_{||}$ and  $q_{\perp}$ are components 
of the spatial frequency parallel and perpendicular to the direction of refractive angle.
The power spectrum $\langle | P_{S,1}(\tau)|^2\rangle_s$ can also be written as 
\begin{equation}
\begin{array}{l}
{ \langle| P_{S,1}(\tau)|^2\rangle_s=
\begin{cases}
H^2_0m^2_1\left(\frac{1}{\tau_{Fr,1}}\right),
&\rm{for}\ \tau<\tau_{Fr,1} \\
H^2_0m^2_1(\left(\frac{1}{\tau_{Fr,1}}\right)\left(\frac{\tau_{Fr,1}}{\tau}\right)^{\alpha_1+1},
&\rm{for}\ \tau>\tau_{Fr,1}
\end{cases} }
\\
\mathrm{where\ } 
\tau_{Fr,1}=z_1\theta_0/(\pi\nu_0\rho_{Fr,1})
\end{array}
\label{q34}
\end{equation}

For the second component, $\langle| P_{S,2}(\tau)|^2\rangle_s$, we find that
\begin{equation}
\begin{array}{l}
\langle| P_{S,2}(\tau)|^2\rangle_s
\begin{cases}
= H^2_0m^2_2\left(\frac{1}{\tau_{Fr,2}}\right),&
\rm{for} \ \tau<\tau_{Fr,2}\\
\approx H^2_0m^2_2\left(\frac{1}{\tau_{Fr,2}}\right)\left(\frac{\tau_{Fr,2}}{\tau}\right)^{\alpha_2+1}, &
\rm{for} \ \tau>\tau_{Fr,2} 
\end{cases} 
\\
\mathrm{where\ } \tau_{Fr,2}=z_2 \theta_0/(\pi\nu_0\rho_{Fr,2})
\end{array}
\label{q35}
\end{equation}
Note that the function $P_S(\tau)$ is a random function of $\tau$. 
Hence, if $\tau > \tau_{Fr,1},\,\tau_{Fr,2}$, 
the distribution of $P_S(\tau)$ is approximately normal for fixed $\tau$. Then, for a normally 
distributed random complex value, we have the relationship
\begin{eqnarray}
\langle | P_S(\tau) | \rangle_s = \frac{\sqrt{\pi}}{2}(\langle | P_S(\tau)|^2 \rangle_s)^{1/2}.
\end{eqnarray}
If $m_1 \approx m_2$, then $\langle | P_{S,1}(\tau)|^2 \rangle_s$ will be 
the primary contributor to \mbox{$\langle | P_S(\tau)|^2 \rangle_s$} 
for $\tau<\tau_{Fr,2}$, while $\langle | P_{S,2}(\tau)|^2 \rangle_s$ 
will affect $\langle | P_S(\tau)|^2 \rangle_s$ predominantly for $\tau>\tau_{Fr,2}$.  

\section[]{Observations and initial data reduction}
\label{obs}

We observed the pulsar \object{B0950+08} for one hour on January 25, 2012 using the RadioAstron 10-m space radio telescope in 
concert with the Arecibo 300-m telescope  and the Westerbork synthesis array. We observed dual polarizations across a 16 MHz band centered on 324 MHz. 
Data were recorded continuously for 5 min scans with a 30 s interval after each scan to write the data to disk. 

Using the Astro Space Center correlator, we performed the first steps of data reduction, which involved removing the dispersion from the pulsar signal, calculating the complex spectrum for each telescope, and calculating the cross-spectra 
for all pairs of telescopes.
The signal was correlated in a 15 ms gate around the maximum of the average pulse, and the noise was evaluated in a gate separated by 52 ms from the maximum. 
For most of our analysis, we averaged the correlator output over four pulsar periods (${\sim}1$ s) and employed a frequency resolution of 125 kHz (128 channels);
although we also used single-pulse spectra with the same frequency resolution in some cases, as noted below.


\section{Analysis and Results}\label{sec:analysis_and_results}
\subsection{Spectra and Correlation Analysis}
\label{sub4.1}

Pulsar \object{B0950+08} has a high level of intrinsic variability, and even exhibits giant pulses with more than 100 times the mean 
flux density \citep{sm3}. 
Because this intrinsic variability is much more rapid than that of scintillation, we normalized each spectrum by its mean value.
We also corrected the spectra for the receiver passband, which we estimated by averaging the off-pulse spectra over the entire observation (3570 s).

\begin{figure}
\includegraphics[width=1.0\textwidth,trim=0.0cm 0cm 0cm 0cm,clip]{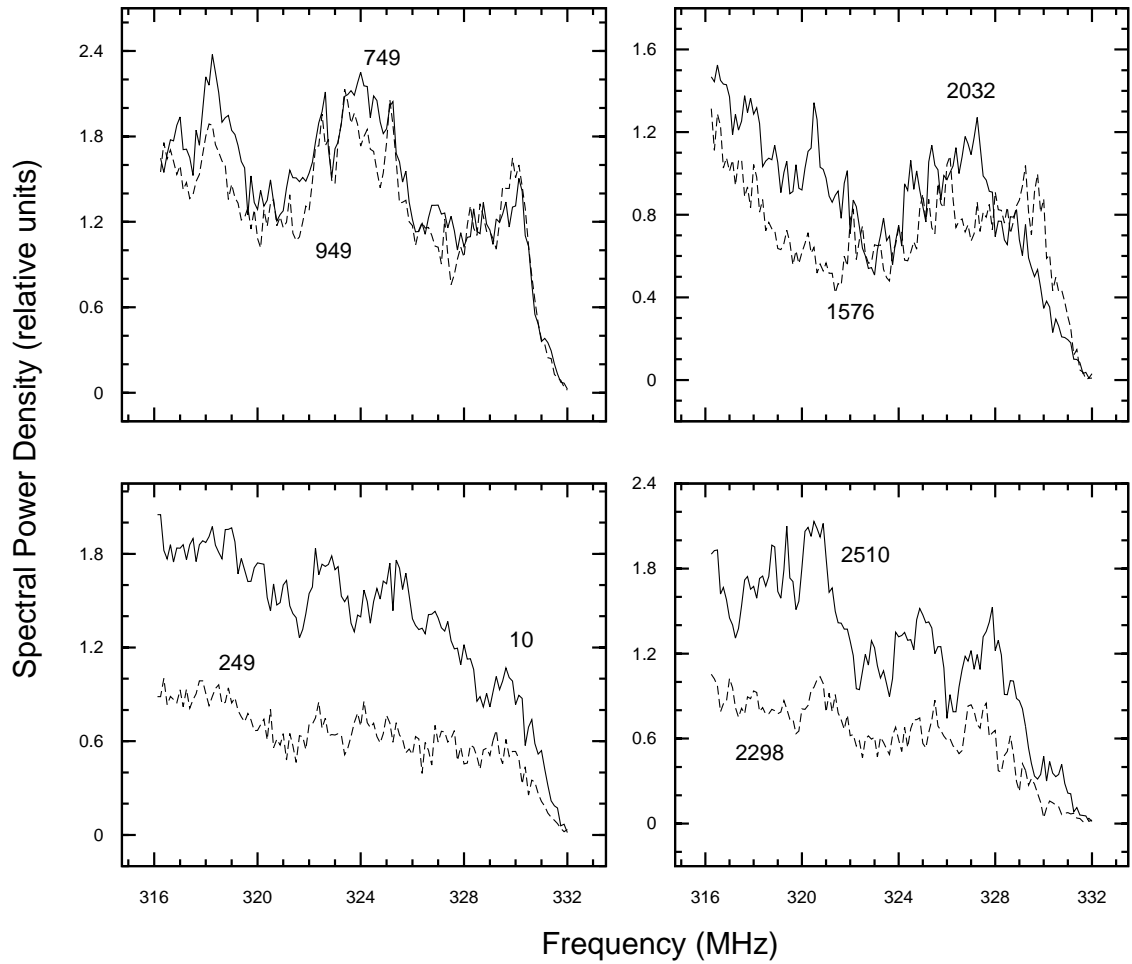} 
 \caption{Spectra of individual pulses for the Arecibo-Westerbok baseline at varying times. The numbers near each plotted curve correspond to the 
 time in seconds from the beginning of the observation. Dashed and solid lines
 correspond to different times.  \label{fig.2}}
\end{figure}

The passbands show narrow-scale interference (only one frequency channel in bandwidth) throughout the entire observation. 
This interference increases signal in individual channels of single-dish spectra, but reduces gain in the corresponding channels of cross-power spectra. 
Hence, for each spectrum, we replaced the intensities in contaminated channels with the mean value of two neighboring channels. 
Figure~\ref{fig.2} shows several individual pulsar cross-power spectra at 
different times for the Arecibo-Westerbork baseline, prior to normalization by the mean intensity. 
Two frequency scales of variability are evident: a small scale, 30 - 40 channels (3.75 - 5) MHz, as well as a large scale, exceeding the 16 MHz receiver band. The small-scale structure changes 
only 
slowly with time, 
and remains 
similar over 
time periods of 
200 s (pulses 749 and 949) and 540 s (pulses 1576 and 2032), whereas the character of the wider structure 
changes little over the 1-hour observation. 

\begin{figure}
\includegraphics[width=1.\textwidth,trim=0.0cm 0cm 0cm .0cm,clip]{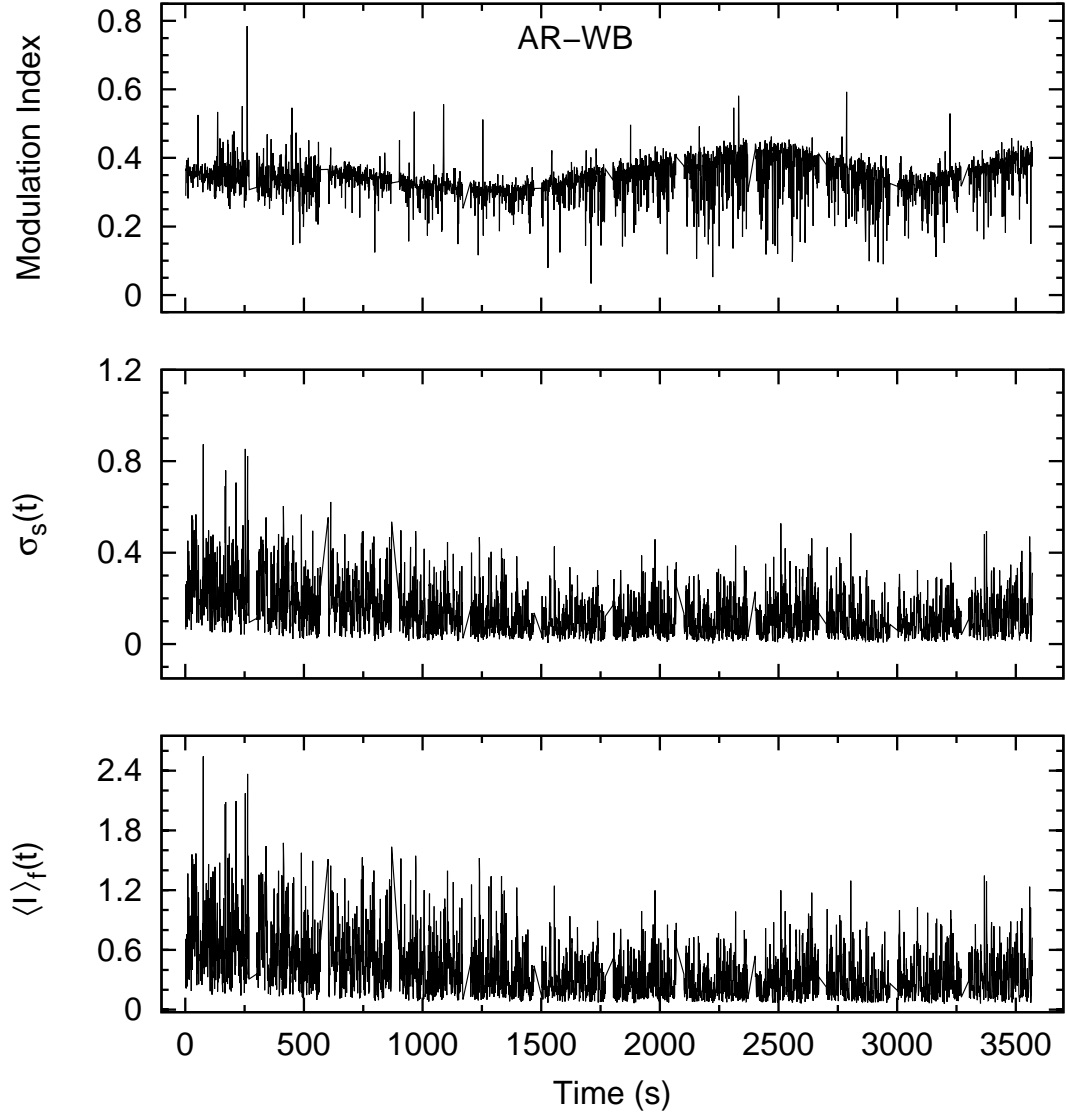} 
 \caption{
Modulation index (top), standard deviation $\sigma$ (middle), and mean value of intensity averaged over frequency for each spectrum (bottom) as a function of time for the Arecibo-Westerbork baseline. 
The time separation between spectra is 1 s. Axis $y$ for middle and bottom pictures is
in arbitrary intensity units.  \label{fig.3}}
\end{figure}

Figure~\ref{fig.3} shows the variability of individual pulses over the experiment.
The lower panel of the figure shows 
the mean intensity $\langle I\rangle_f$ of individual pulses as a function of time. 
Here, the subscripted angular brackets $\langle ... \rangle_f$ indicate an average over frequency channels of one single-pulse integration. 
Fast variability reflects intrinsic pulse-to-pulse varaiations, wheras the slow modulation reflects scintillation. The 
middle panel of Figure~\ref{fig.3} shows the standard deviation of
flux density in individual single-pulse spectral $\sigma_S(t)$ as a function of time. 
For each spectrum, we used the autocorrelation function at a lag of one channel (to exclude the contribution of noise) 
to find $\sigma_{S}(t)$.
In the upper panel of the figure we show the modulation index $m(t)=\sigma_S(t)/\langle I \rangle_f(t)$ as a function of time; null sequences correspond to 30\ s gaps in data. The modulation 
index $m$ is equal to 0.35 to 0.4, indicating weak scintillation at $\nu_0  = 324$ MHz for \object{PSR~B0950+08}. The fast 
fluctuations of $m(t)$ are due to noise or weak pulses. 
The slow variations of $m(t)$ are caused by the wide-bandwidth component of scintillation; the narrow-bandwidth component of scintillation averages out over the observing band.

\begin{figure}
\includegraphics[width=1.\textwidth,trim=0.0cm 0cm 0cm .0cm,clip]{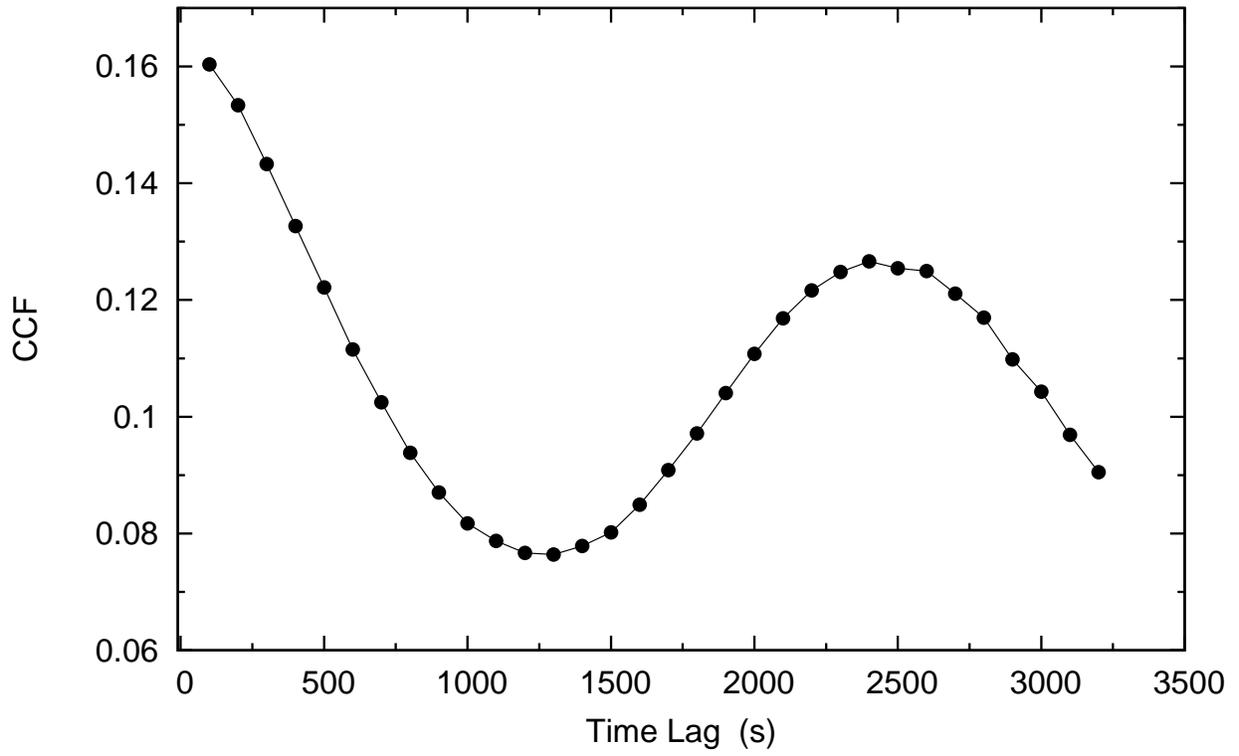} 
 \caption{
The mean value of the cross correlation after averaging over frequency, separated by $100 k$ s, where $k$ = 1, 2, 3, \ldots, 35, for the Arecibo-Westerbork baseline.  \label{fig.4}}
\end{figure}

Figure~\ref{fig.4} shows the mean cross-correlation of the interferometric visibility on the Arecibo-Westerbork baseline,
after the visibility is averaged over frequency:
\begin{eqnarray}
{\rm CCF} (\Delta t) = \big\langle  V(f,t)  V^*(f,t+\Delta t)  \big\rangle_{f,t} 
\label{eq:CCF_def}
\end{eqnarray}
The correlation is shown for nonzero temporal lags of $\Delta t = 100 k$ s, where $k$ = 1, 2, \ldots, 35.  
Intrinsic pulsar fluctuations are uncorrelated at these large lags, so the correlation arises from scintillation. The characteristic timescale of scintillation, defined as the lag at half maximum, is $t_{sc} \approx 1000$ s. Because the correlation is averaged over frequency, this reflects the effect of the broadband variation in Figure\ \ref{fig.3}.

\subsection{Estimated structure function}\label{sec:estimated_structure_function}

We used our observations to estimate the structure function of the interferometric visibility,
and compared this estimate with the theoretical results of Section\ \ref{main}.
To form our estimate, we normalized each spectrum $V(f,t)$ by its mean in frequency, and also corrected for the receiver
bandpass:
\begin{eqnarray}
F(f,t)=\frac{ V(f,t) B_0}{ \langle V \rangle_f(t) B(f)}
\label{eq:Fdef}
\end{eqnarray}
where $B(f)$ is the receiver bandpass and $B_0$ is its value at the center frequency. As mentioned 
in subsection~\ref{sub1.1} and discuss further in Appendix \ref{sec:appendix_a}, we used the squared intensity when the noise level was 
comparable to the signal (i.e., on the space baselines). So we calculated a mean structure function 
for both baselines as: 
\begin{eqnarray}
{\mathcal F}(\Delta f,\Delta t)=\bigg\langle [F^2(f,t)-F^2(f+\Delta f,t+\Delta t)]\cdot 
[F^2(f,t+\Delta t_1)-F^2(f+\Delta f,t+\Delta t+\Delta t_1)]\bigg\rangle_{f,t}
\label{eq:structfuncdef}
\end{eqnarray}
where we include an additional shift in time $\Delta t_1 = 20$ s for a significant reduction of the effects of noise for the 
structure function at zero frequency lag. 
As we show in Appendix \ref{sec:appendix_a}, in weak scintillation the structure function for the squared modulus of the visibility 
is proportional to the structure function for the modulus of visibility. Consequently, for 
our calculated normalized structure function we can use all theoretical relations from section~\ref{main}. 
Structure functions for Arecibo-Westerbork and RadioAstron-Arecibo baselines normalized by $(\langle F^2(f,t)\rangle_{f,t})^2$ 
are plotted with frequency lag $\Delta f$ in Figure~\ref{fig.5}, for different time lags $\Delta t$. We calculated the structure function for both positive and 
negative frequency lags, $\Delta f$.

\subsection{Comparison with theory}\label{sec:comparison_with_theory}

\subsubsection{Simple model}\label{sec:simple_model}

\begin{figure}
\includegraphics[width=1.\textwidth,trim=0.0cm 0cm 0cm .0cm,clip]{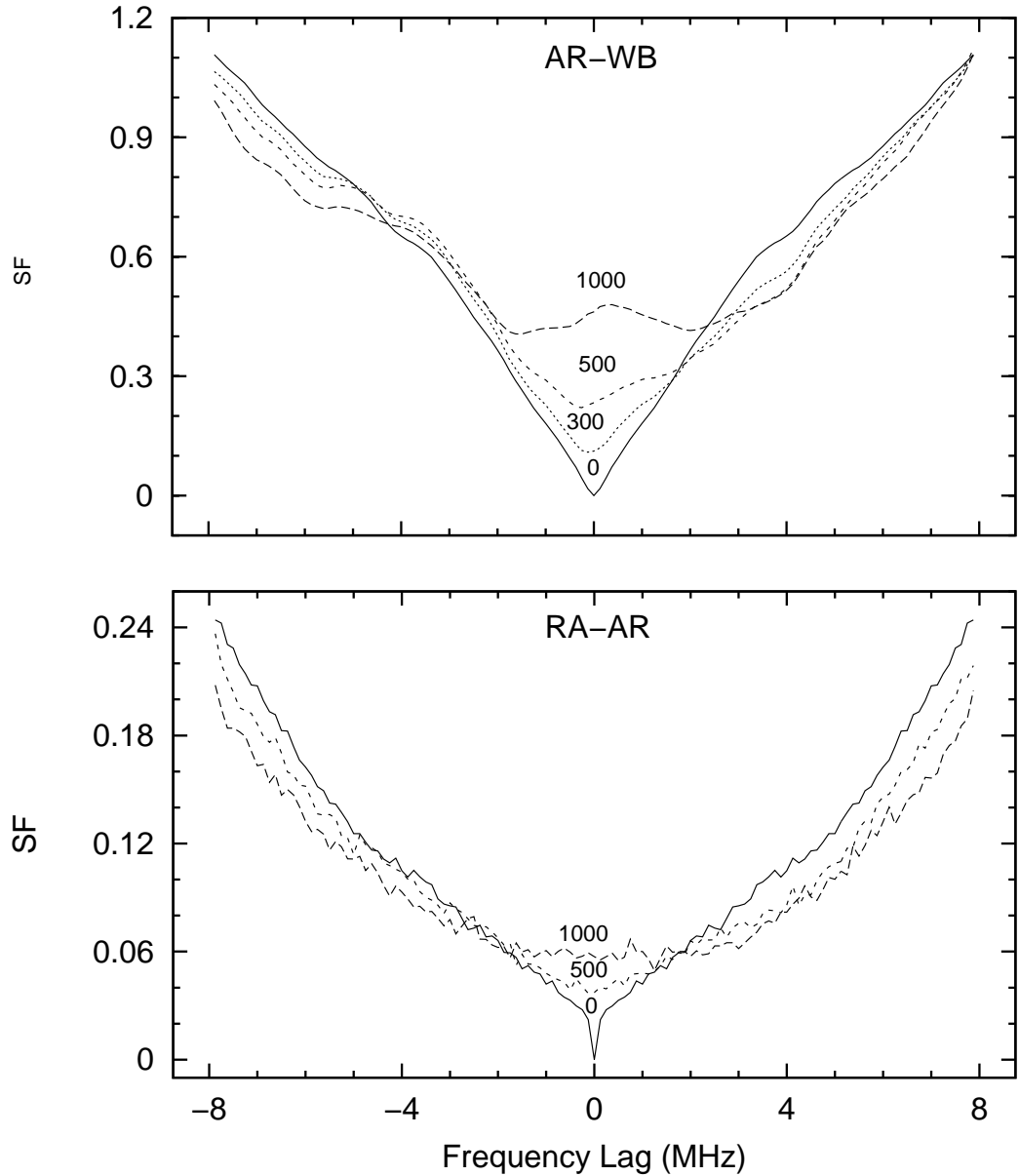} 
 \caption{Mean structure functions for different time lags $\Delta t$ on the Arecibo-Westerbork 
 baseline (upper figure)  and  on the RadioAstron-Arecibo 
 baseline (lower figure). 
 The numbers in the figure correspond to $\Delta t$ in s.
  The narrow feature at $\Delta f=0, \Delta t=0$ is due to noise.
The $y$-axes have the same units as in Figure \ref{fig.8}.
  \label{fig.5}}
\end{figure}

The structure functions shown in the two panels of Figure\ \ref{fig.5}  have qualitatively different forms.
Comparison of the two shows that the structure function on the shorter Arecibo-Westerbork baseline comprises a narrow-bandwidth component
and a broader-bandwidth component.
For the long 
RadioAstron-Arecibo space baseline, the narrow-bandwidth component is absent; we see only the broad-bandwidth structure. 
(The sharp detail 
at lag $\Delta t = 0$ s is caused by noise.)
The narrower component also appears only at small time lags, whereas the broader component appears at both large and small time lags.
The two frequency scales correspond to two effective layers of turbulent 
plasma, separated in space, where scattering of pulsar emission take place.  

\begin{figure}
\includegraphics[width=1.\textwidth,width=12.cm]{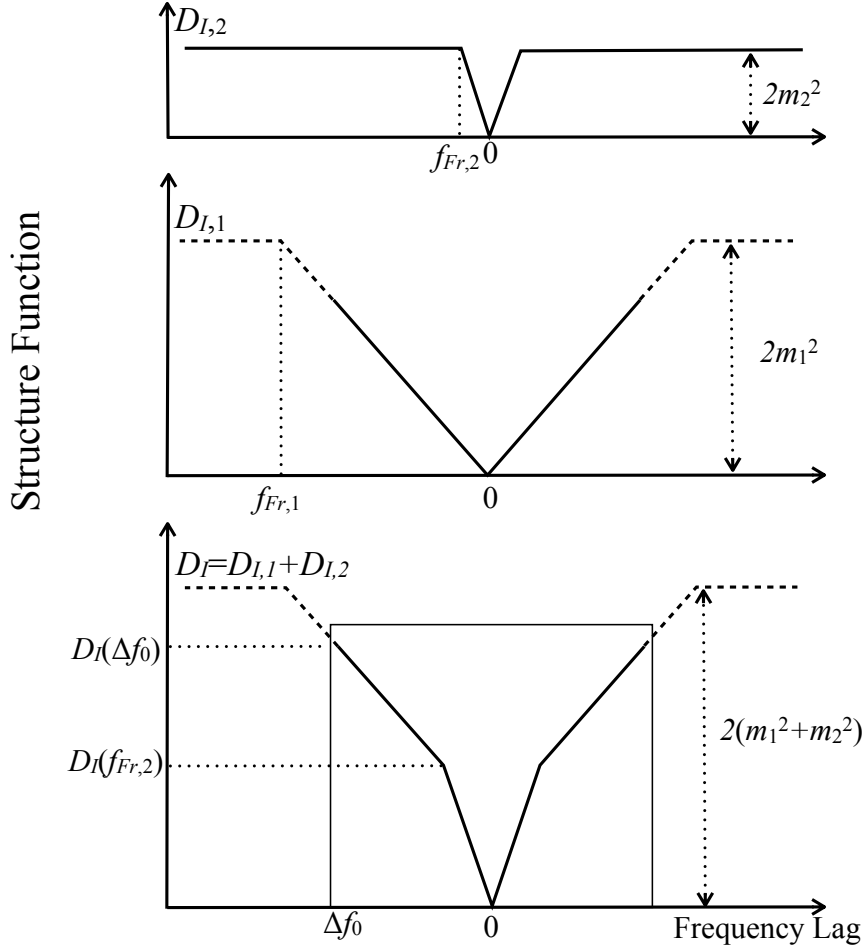} 
\caption{
Scheme of dissection of structure function into effects of near and far screens.
Upper: Structure function for distant screen $D_{I,2}(\Delta f)$, plotted with frequency lag $\Delta f$.
Middle: Structure function for nearby screen $D_{I,1}(\Delta f)$.
Lower: Sum of structure functions for the two screens $D_{I}(\Delta f)$, modeling observations on the Arecibo-Westerbork baseline.
Box shows area of Figure \ref{fig.5}.
\label{fig:schematic}}
\end{figure}

\begin{figure}
\includegraphics[width=1.0\textwidth,trim=0cm 0cm 0cm .0cm,clip]{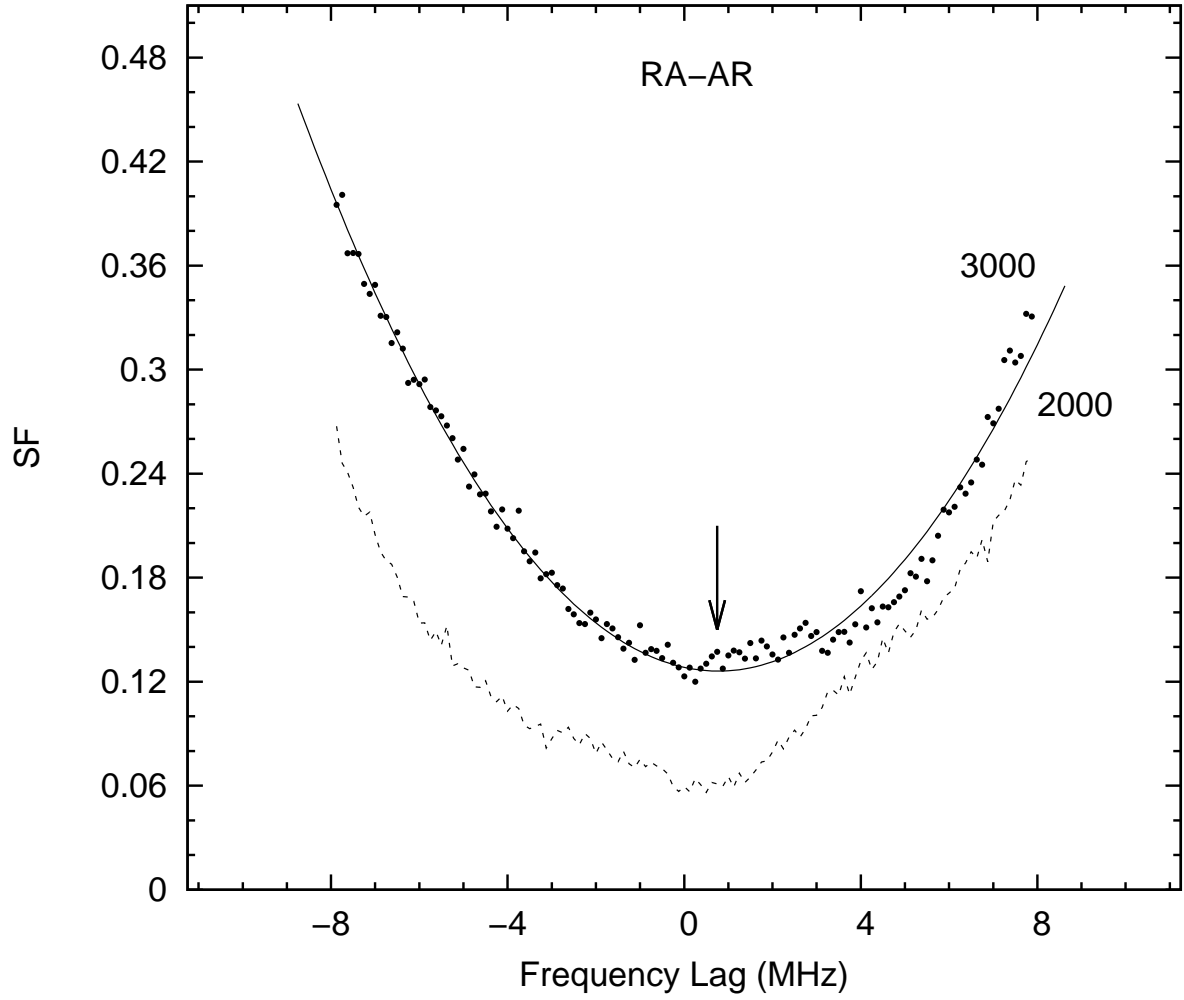} 
\caption{
The mean structure function for the RadioAstron-Arecibo baseline for time lags 2000 s and 3000 s. A parabola was fitted to the structure  function for time lag 3000 s (line). 
The minimum of the fitted parabola is marked by an arrow. 
\label{fig.8}}
\end{figure}

As a simple model for the structure function, we adopt a piecewise-linear form,
displayed in Figure\ \ref{fig:schematic}.
Formally, we take $\alpha_1=\alpha_2= 1$ in Equation\ (\ref{eq:phase_structure_fns});
this is adequate for the determination of characteristic scales.
This leads to the form
\begin{eqnarray}
D_{I,\ell}({\mathbf r}_\ell) &=& 
\begin{cases}
m^2 \frac{ | {\mathbf r}_\ell | }{\rho_{Fr,\ell}} , & |{\mathbf r}_\ell | < {\rho_{Fr,\ell}} \\
m^2 , & | {\mathbf r}_\ell | \ge {\rho_{Fr,\ell}}
\end{cases} \\
{\rm where}\quad\quad  \frac{ | {\mathbf r}_\ell | }{\rho_{Fr,\ell}} &=& 
\left| \frac{ \Delta{\boldsymbol\rho} }{\rho_{Fr,\ell}} + \frac{ \Delta t}{t_{Fr,\ell}} + \frac{{\boldsymbol\theta}_0 \Delta f}{f_{Fr,\ell}} \right|
\label{eq:relldef}
\end{eqnarray}
where we have combined results of Sections \ref{sec:time_freq_baseline} through \ref{sec:Fresnel_scales_freq_time}.  
The index $\ell=1,2$ runs over the two screens.
Note that the displacements are added vectorially to form the arguments ${\mathbf r}_\ell$, but the arguments for different screens $\ell$ are completely
independent.

Although the dependence of the structure function on $|{\mathbf r}_\ell |$ is linear,
interplay among the arguments can lead to different dependences on $\Delta f$.
For example, if 
dispersion by the cosmic prism is perpendicular to the velocity of the ray through the screen so that $\beta_\ell \approx \pi/2$,
and both frequency and time offsets contribute with $\Delta f/f_{Fr} < 1$ and $|{\mathbf V} \Delta t | / r_{Fr} < 1$,
then the variation of $D_{I}$ with $f$ becomes approximately quadratic, as Equation\ (\ref{eq:beta_phi_def}) shows.
Likewise, for $\Delta f/f_{Fr}< |\Delta {\boldsymbol\rho}| / r_{Fr}< 1$ and $\varphi\approx \pi/2$,
the dependence is approximately quadratic.

On the space-Earth RadioAstron-Arecibo baseline,
the long baseline suppresses most of the narrowband structure and some of the wideband structure,
as Figure \ref{fig.5}~(bottom) shows.  The shape of the structure function becomes more quadratic with increasing $\Delta t$.
As we argue in more detail below, this suggests that the interferometer baseline is perpendicular to the velocity of the ray relative to the more distant screen, screen 2.

The two components of the structure function on the shorter baseline represent the structure functions from the two screens:
\begin{eqnarray*}
D_{I,AR-WB} &\equiv& D_I=D_{I,1}+D_{I,2}
\end{eqnarray*}
Consequently, for the shorter Arecibo-Westerbork baseline, 
we expect the structure function to take the form:
\begin{eqnarray}
D_{I}(\Delta f)=
\begin{cases}
2m^2_1 \left| \frac{\Delta f}{f_{Fr,1}}\right|+2m^2_2\left|  \frac{\Delta f}{f_{Fr,2}}\right|, 
&{\rm for\ }  | \Delta f | <f_{Fr,2} \\
2m^2_1\left| \frac{\Delta f}{f_{Fr,1}}\right|+2m^2_2 
&{\rm for\ }  f_{Fr,1} > | \Delta f | >f_{Fr,2} \\
2m^2_1+2m^2_2 
&{\rm for\ }   | \Delta f | >f_{Fr,1}
\end{cases}
\label{q42}
\end{eqnarray}
The increase of the structure function up to the maximum frequency difference that is reliably sampled for our 16-MHz observing bandwidth,
$\Delta f_0 = 8$ MHz, indicates that $f_{Fr,1}$ is equal to or greater than $\Delta f_0$. 
This increase appears for both long and short baselines.
Consequently, the present data do not allow us to explore the third possibility, $| \Delta f | >f_{Fr,1}$, in Equation\ (\ref{q42}).
From inspection of Figure~\ref{fig.5}~(top), the structure function on the shorter baseline has a change of slope at the frequency corresponding to the 
Fresnel scale for screen 2,  at $\Delta f =  f_{Fr,2} = 3.1$ MHz (25 channels).
This is the transition from the first possibility to the second in Equation\ (\ref{q42}).

\subsubsection{Evaluation of model parameters}

From inspection of Figure\ \ref{fig.5} we observe that $D_I$ approximately doubles between $f_{Fr,2}=3.1\ {\rm MHz}$ and $\Delta f_0= 8\ {\rm MHz}$:
\begin{eqnarray}
D_I(f_{Fr,2}) \approx 0.5 D_I(\Delta f_0) 
\end{eqnarray}
If we substitute for $D_I$ into this expression
from Equation\ (\ref{q42}), using second possibility because $f_{Fr,1}>\Delta f_0 > f_{Fr,2}$,
we find:
\begin{eqnarray}
2 m_1^2 \frac{f_{Fr,2}}{f_{Fr,1}}+ 2 m_2^2 &\approx& 0.5 \left( 2 m_1^2 \frac{\Delta f_0}{f_{Fr,1}}+ 2 m_2^2 \right)
\label{eq:D_I_relation_consts}
\end{eqnarray}
%
Note also that the modulation index found for the entire scan, $m=0.35$, as illustrated in Figure\ \ref{fig.3},
is approximately equal to $D_I(\Delta f_0)$, because the bandwidth of the observations limits the modulation index.
Thus:
\begin{eqnarray}
D_I(\Delta f_0) = 2 m_1^2 \frac{\Delta f_0}{f_{Fr,1}}+ 2 m_2^2 \approx 2 m^2 
\label{eq:m_and_D_I}
\end{eqnarray}
From Equations (\ref{eq:D_I_relation_consts}) and (\ref{eq:m_and_D_I}), we find that 
\begin{eqnarray}
m_1 &=& \sqrt{\frac{f_{Fr,1}}{2 \Delta f_0 + 2 f_{Fr,2}} } m \label{eq:m1_expression}\\
m_2 &=& \sqrt{\frac{\Delta f_0 - 2 f_{Fr,2}}{ 2 \Delta f_0 -2 f_{Fr,2}}}  m = 0.15
\end{eqnarray}
We can set bounds on $m_1$ from the facts that $f_{Fr,1}>\Delta f_0 = 8\ {\rm MHz}$,
as observed above;
and from the fact that \citet[][see Figure~10]{sm2} find $f_{Fr,1}<15\ {\rm MHz}$,
at our observing frequency.
We then find
\begin{eqnarray}
0.32 < m_1 < 0.43
\end{eqnarray}
The smaller value corresponds to our lower limit on $f_{Fr,1}$, and the larger to the upper limit.
Consequently, our model comprises two screens, both weakly scattering.
Screen 1 has the greater modulation index $m_1$, and a larger frequency scale corresponding to the Fresnel scale $f_{Fr,1}$.
This suggests that screen 1 is closer,
as can be seen from Equations\ (\ref{eq:fFr1_def} and\ \ref{eq:fFr2_def}).

\subsubsection{Scales and distances}\label{sec:scales_and_distances}

%
We can use the behavior of the structure function with time, as well as frequency,
to estimate the Fresnel scales and distances of the scattering screens.
The amplitude of the narrower component of the structure function 
decreases with increasing of time shift, and falls to zero at time lag,  $\Delta t = 1000$ s. 
The cross-correlation coefficient of spectra decreases with time lag and falls to half-maximum as discussed in Section\ \ref{sub4.1} above, and as shown in 
Figure~\ref{fig.4}.
This suggests that the typical timescale for the narrow component, produced by screen 2, is less than 1000 sec.

We can evaluate the Fresnel scale for screen 1 by comparing the structure function at $\Delta f=0$ and $\Delta f=\Delta f_0$,
at $\Delta t=1000\ {\rm s}$.
At this time lag, the structure function at $\Delta f=0$ is 0.42 times that at $\Delta f=\Delta f_0$ as seen in Figure~\ref{fig.5}.
\begin{eqnarray}
D_{I}( \Delta f=0, \Delta t=10^3\ {\rm s}) = 0.42 D_{I}( \Delta f=\Delta f_0, \Delta t=0) 
\label{eq:compare_DIs}
\end{eqnarray}
Using the forms for the structure functions introduced in Section\ \ref{sec:simple_model}, and observing that the contribution of $D_{I,2} = 2 m_2^2$
at this large time lag, Equation\ (\ref{eq:compare_DIs}) becomes:
\begin{eqnarray}
\left( \frac{V_1\,\Delta t}{\rho_{Fr,1}}\right) m_1^2 + 2 m_2^2  =0.42 \left[ 2 \left( \frac{\Delta f_0}{f_{Fr,1}}\right) m_1^2 + 2 m_2^2 \right] 
\end{eqnarray}
From this we obtain an expression for $\rho_{Fr,1}$:
\begin{eqnarray}
\rho_{Fr,1}=\frac{m_1^2 f_{Fr,1}  V_1 \Delta t}{0.42 \, \Delta f_0 \, m_1^2 - (1-0.42) f_{Fr,1} m_2^2} 
\end{eqnarray}
We eliminate $m_1$ in favor of $m$ and $f_{Fr,1}$ using Equation\ \ref{eq:m1_expression} and find:
\begin{eqnarray}
\rho_{Fr,1}&=& (1.4{\rm\ to\ }2.7)\times 10^{5}\ {\rm km}
\end{eqnarray}
where in the last line we have used our observational limits for $f_{Fr,1}$ $8\ {\rm MHz} < f_{Fr,1} <15\ {\rm MHz}$, and for $V_1$ used the 
speed of the Earth, relative to the Local Standard of Rest, at the date of observation $V_1=V_{\
obs}=41\ {\rm km\ s}^{-1}$.
Using Equation\ (\ref{eq:Fresnel_close}) we find for the distance of screen 1, $z_1$,
\begin{eqnarray}
z_1= k(\rho_{Fr,1})^2=(4.4 - 16.4) \quad \rm{pc}
\end{eqnarray}
Thus, screen 1 is quite close to the Earth.

Using Equation\ (\ref{eq:fFr1_def}), 
we obtain for the refractive angle
\begin{eqnarray}
\theta_0 &=&\frac{\rho_{Fr,1}}{2 z_1}\frac{\nu_0}{f_{Fr,1}} \label{eq:theta_0_def} \\
&=& (1.1 - 4.4) \quad \rm{mas} \label{eq:theta_0_value}
\end{eqnarray}
where the larger value for $\theta_0$ arises from the smaller value for $f_{Fr,1}$, leading to a closer screen (smaller value of $z_1$).

Using this value for refractive angle and $f_{Fr,2}  = 3.1$ MHz in the 
expression for the frequency scale corresponding to refraction by the nearby screen, Equation\ (\ref{eq:fFr2_def}),
we find for the value of $z_2$
\begin{eqnarray}
z_2 = (26 - 170) \quad \rm{pc}
\end{eqnarray}
where the range of screen distances arises from the range of possible values for $f_{Fr,1}$,
with the closer distance for screen 2 associated with a closer distance for screen 1, and both arising from the smaller value for $f_{Fr,1}$.
Screen 2 is  always more distant, and may lie at a significant fraction of the pulsar distance of $260\ {\rm pc}$.

The Fresnel scale for screen 2 is given by Equation\ (\ref{eq:Fresnel_far}):
\begin{eqnarray}
\rho_{Fr,2}=\left(3.5{\rm\ to\ }15\right)\times 10^{5}\ \rm{km}
\end{eqnarray}
where the smaller value corresponds to the lower limit on $f_{Fr,1}$, and the larger to the upper limit.


We note that our reconstruction of Fresnel scales, screen distances, and refraction angle is roughly consistent with the observation that the narrow component of the structure function is supressed on the long
RadioAstron-Arecibo baseline, as shown in Figure\ \ref{fig.5}~(bottom).
We expect this component to become decorrelated over a distance of $\rho_{Fr,1}$, or 140,000 to 270,000\ km, whereas the projected baseline length is 220,000 km.
Our interferometer results favor the greater distances in this range from screen 1.

%

\subsubsection{Asymmetry of the structure function}\label{sec:asymmetry}

The cosmic prism disperses the scintillation pattern across the observer plane,
so that particular intensity maxima and minima appear at different positions at different frequencies.
If the screen moves parallel to the direction of dispersion, then the observer notes a shift in the frequency of the scintillation pattern with time, as given by 
${\mathbf V}_{1}$ and ${\mathbf V}_{2}$ in Equations\ (\ref{q15}-\ref{eq:time_shift_2}).
If multiple screens are present, as we argue above, and if they have different velocities, then the observer will notice different rates
for the different resulting patterns.

The shift in frequency of the scintillation pattern with time or position leads to an asymmetry in frequency $\Delta f$ of the structure function:
$D_{I,\ell}( \Delta f , \Delta t, \Delta \rho )$, 
for nonzero time lag $\Delta t$ or finite baseline length $\Delta \rho$. 
This asymmetry increases proportionately with $\Delta t$ or $\Delta \rho$.
For two screens with different velocities, 
displacements of the structure functions $D_{I,1}$ and $D_{I,2}$ increase with different rates.
If the screens are moving with velocities with opposite directions of projection onto ${\boldsymbol\theta}_0$,
the resulting displacements have opposite sign.

As an example of this asymmetry, Figure~\ref{fig.8} shows
the mean structure functions for the long RadioAstron-Arecibo baseline at large time lags. The line shows the best-fitting parabola at $\Delta t = 3000$ s. The minimum is shifted by 750 kHz, or 6 channels, toward $+\Delta f$. As Equation\ (\ref{eq:r_def}) suggests, a minimum of the structure function lies where 
\begin{equation}
\rho_f=-\rho_t
\end{equation}
$\rho_t=V_1\cos\beta_1\Delta t=1.8 \cdot 10^9$ cm. Using Equations~(\ref{eq:fFr1_def}, \ref{eq:m_and_D_I}, and \ref{eq:theta_0_value}) we find: 
$z_1\theta_0=\rho_f\cdot\nu_0/(2\Delta f)$. From the previous expressions using $z_1 = 4.4 {\rm\ to\ } 16.4$ pc we find the refractive angle is
$\theta_0 = 1.4 {\rm\ to\ } 5.8$ mas, which is in reasonable agreement with the value obtained previously. 

We can estimate directions of refractive dispersion and screen velocity
from the measured asymmetry of the structure function.
For our observations on the short Arecibo-Westerbork baseline, 
the weak asymmetry of the structure function, along with a relatively rapid decorrelation of the narrow component at $\Delta t = 1000$ s
as seen in Figure\ \ref{fig.5} (top),
suggests that 
the angle $\beta_1$ between the vectors ${\boldsymbol\theta}_0$ and ${\mathbf V}_{obs}$ 
is close to $\pi/2$. 

We quantify the asymmetry of the normalized difference of structure functions for positive and negative frequency lags by the ratio: 
\begin{equation}
{\mathcal D}(\Delta f, \Delta t) = \frac{D_{I}(\Delta f,\Delta t)- D_{I}(-\Delta f,\Delta t)}{D_{I}(\Delta f,\Delta t)+D_{I}(-\Delta f,\Delta t)}
\label{eq:asymmetry_ratio}
\end{equation}
For particular values of $\Delta t$, the extrema as a function of $\Delta f$ lie at
\begin{equation}
\partial_{\Delta f} {\mathcal D}(\Delta f, \Delta t) = 0
\label{eq:extremal_condition}
\end{equation}
For weak-scattering models such as the simple model discussed in Section\ \ref{sec:simple_model},
where the structure function for each screen has a single minimum
and becomes constant for large displacements,
extrema of ${\mathcal D}$ tend to lie near or at the minima of the structure functions of the individual screens.

We display the ratio ${\mathcal D}(\Delta f, \Delta t)$ for our observations in Figure\ \ref{fig.7},
at a time lag of $\Delta t = 1000$ s, for both RadioAstron-Arecibo and Arecibo-Westerbork baselines. 
For both, the asymmetry is relatively mild: $|{\mathcal D}| \ll1$.

For the Arecibo-Westerbork baseline, we have two extrema on the  curve along the positive x-axis $\Delta f>0$.
A minimum lies at large values  $\Delta f_1 \approx 3\ {\rm MHz}$,
and a maximum lies at a small values of $\Delta f_2 \approx 1\ {\rm MHz}$. 
The presence of both a maximum and a minimum suggests that the transverse velocities of the two screens 
have opposite projections onto the direction of dispersion, so that they migrate toward opposite directions in frequency with increasing time.
For this relatively short baseline, $\Delta{\boldsymbol\rho}_{AW}\approx 0$,
as Figure \ref{fig:vectors} suggests.

On the long RadioAstron-Arecibo baseline, we see only a single minimum at a position close to $\Delta f_1$.
Under the assumption that the short baseline is sensitive to both broadband scintillation (from the nearby screen 1)
and narrowband scintillation (from the distant screen 2),
whereas the long baseline is sensitive to only the broadband contribution,
we suggest that the minimum at larger frequency difference $\Delta f_1$ is associated
with screen 1, and the maximum at small difference $\Delta f_2$ with screen 2.

\subsubsection{Directions of ray motion relative to screens}

\begin{figure}
\includegraphics[width=1.0\textwidth,trim=0cm 0cm 0cm .0cm,clip]{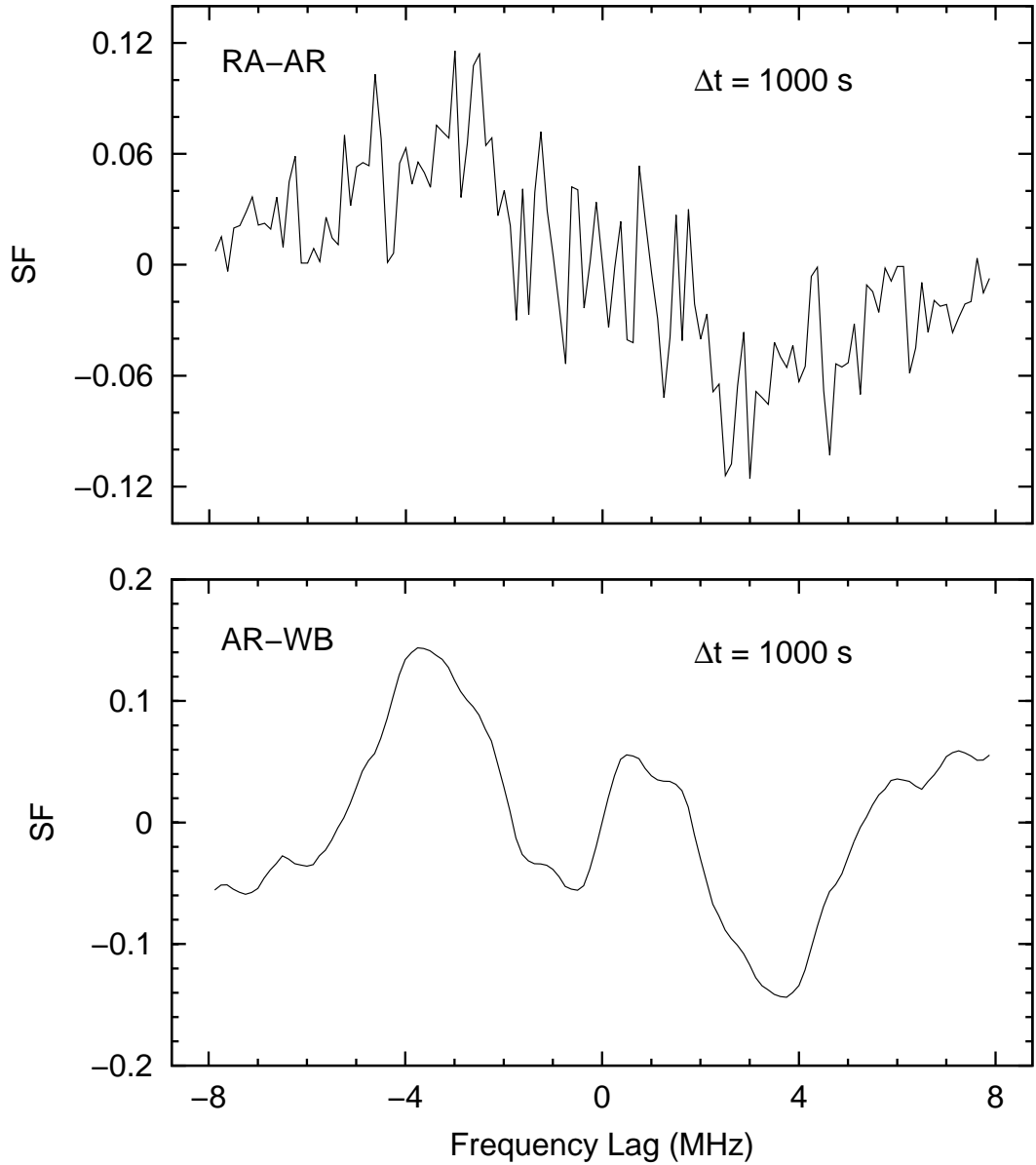} 
\caption{
 The ratio of structure functions difference for positive and negative frequency lags 
on its sum in dependence of frequency lag: for baseline RadioAstron-Arecibo (top), 
for Arecibo-Westerbork (bottom). The time lag is 1000 s.
\label{fig.7}}
\end{figure}

\begin{figure}
\includegraphics[width=1.\textwidth,angle=270,width=15.cm]{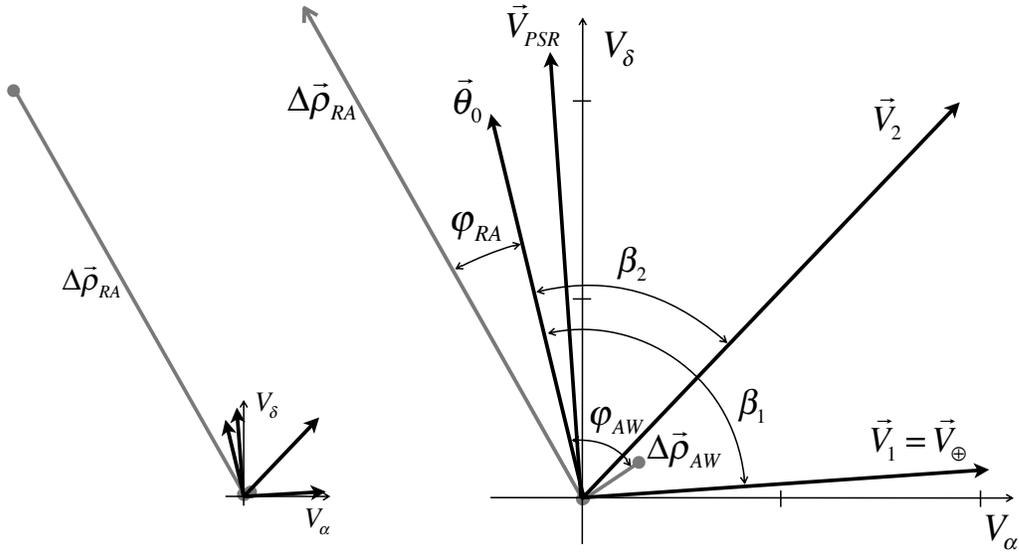} 
 \caption{
Vectors on the sky showing velocity of the pulsar ${\mathbf V}_{PSR}$,
direction of gradient of refracting wedge ${\boldsymbol\theta}_0$,
velocity of the Earth at this epoch, ${\mathbf V}_{obs}={\mathbf V}_{1}$,
velocity of ray relative to screen 2, ${\mathbf V}_{2}$,
and interferometer baselines $\Delta {\boldsymbol\rho}_{RA}$, $\Delta {\boldsymbol\rho}_{AW}$ 
expressed in velocity units: $\Delta{\boldsymbol\rho}/10^3\ {\rm s}$. 
Length of ${\boldsymbol\theta}_0$ is arbitrary.
Left figure shows all vectors,
right figure is enlarged five times.  \label{fig:vectors}}
\end{figure}

By evaluating Equation\ (\ref{eq:asymmetry_ratio})
for the geometry and simple form of the structure function discussed above, 
for 
$\rho_{Fr,2}/{V}_2 <\Delta t <\rho_{Fr,1}/{V}_1$ as in Figure\ \ref{fig.7},
we can evaluate the asymmetry function for the short Arecibo-Westerbork baseline $|\Delta \mathbf{\rho}_{AW}| <\rho_{Fr,2}$.
We find that
the asymmetry function reaches a maximum value of:
\begin{eqnarray}
{\mathcal D}(\Delta f, \Delta t) &\approx & 
\left(\frac{m_2}{m_1}\right)^2\left(\frac{ \rho_{Fr,1}}{\rho_{Fr,2}}\right)\cos \beta_2 \label{eq:asymetry_func_small_max} \\
&\approx& 0.05 {\rm\ to\ }0.1\cos\beta_2 \nonumber
\end{eqnarray}
As the lower panel of Figure\ \ref{fig.7} indicates,
the observed value of ${\mathcal D}$ at the maximum at $\Delta f_2$ is 
\begin{equation}
{\mathcal D}(\Delta f_2, \Delta t = 10^3\ \rm{s}) \approx 0.05 .
\label{eq:asymmetry_value_max}
\end{equation}
Combining Equations\ (\ref{eq:asymetry_func_small_max} and \ref{eq:asymmetry_value_max})
and evaluating the expression for our limits on $\Delta f_{Fr,1}$ 
yields the angle 
\begin{equation}
0 \le \beta_2\le 60^{\circ}. 
\end{equation}
This is the angle between vectors ${\boldsymbol\theta}_0$ and ${\mathbf V}_2$. 

Similarly, for our model, the minimum of the asymmetry function
has the value
\begin{equation}
{\mathcal D}(\Delta f, \Delta t) \approx   \cos \beta_1
\end{equation}
In this case, as the lower panel of Figure\ \ref{fig.7} indicates,
\begin{eqnarray}
{\mathcal D}(\Delta f_1, \Delta t =10^3\ \rm{s})  &\approx& -0.15 \\
\beta_1&\approx& 100^{\circ}
\end{eqnarray}
The angle 
$\beta_1$ is the angle between vectors ${\boldsymbol\theta}_0$ and ${\mathbf V}_1$. 

For the space-Earth baseline 
with length $| \Delta{\boldsymbol\rho}_{RA}|=220,000\ {\rm km}$, the value of the asymmetry function
at its minimum is given by the relation:
\begin{eqnarray}
{\mathcal D}(\Delta f, \Delta t) &\approx& 
({\mathbf V}_{obs}\Delta t/\Delta\rho)\cos\beta_1\approx 0.2 \cos\beta_1 
\end{eqnarray}
The upper panel of Figure\ \ref{fig.7} indicates:
\begin{equation}
{\mathcal D}(\Delta f_1, \Delta t = 10^3\ \rm{s})  \approx -0.04
\end{equation}
The resulting estimated value $\cos\beta_1\approx -0.2$ is in approximate agreement with that obtained from the short baseline.

Figure\ \ref{fig:vectors} summarizes the velocities and the refractive angle that we obtain,
and compares them with the baseline vectors for our time lag of $\Delta t=10^3$\ s and the pulsar velocity.
The difference of the angles $\beta_1$ and $\beta_2$ indicates the presence of two spatially-separated scattering screens.
We assume that ${\mathbf V}_2$ is set by the velocities of the observer ${\mathbf V}_{obs}$ and of the pulsar ${\mathbf V}_{PSR}$;
and set the distance of the screen to $z_2 = 0.5\, z$. This leads to $\beta_2 = 55^{\circ}$.
For screen distances of up to 170 pc, the leveraged pulsar velocity can increase the declination component of ${\mathbf V}_2$ 
to as much as $70\ {\rm km}\ {s}^{-1}$.
The baseline vectors are displayed as velocities, with the length of the baseline divided by $10^3$\ s.

\subsection{Coherence function}\label{ref:sec_coherence_function_measurement}

\begin{figure}
\includegraphics[width=1.\textwidth,trim=0.cm 0cm 0cm 2.0cm,clip]{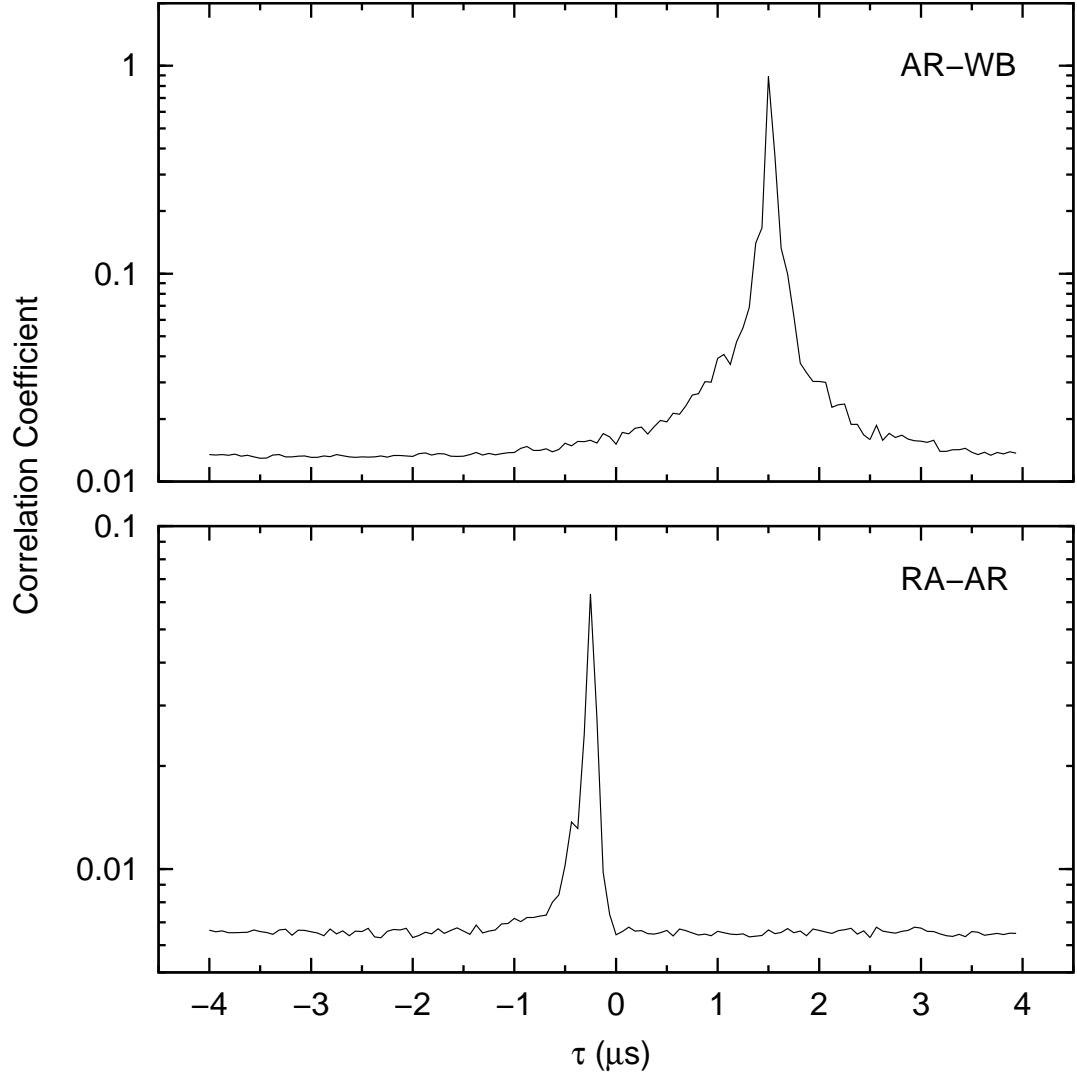} 
 \caption{
The mean coherence function: for the Arecibo-Westerbork (top), and RadioAstron-Arecibo baselines (bottom). The y-axis is amplitude, plotted on a log scale.
\label{fig.9}}
\end{figure}

\begin{figure}
\includegraphics[width=1.\textwidth,trim=0.0cm 0cm 0cm .2cm,clip]{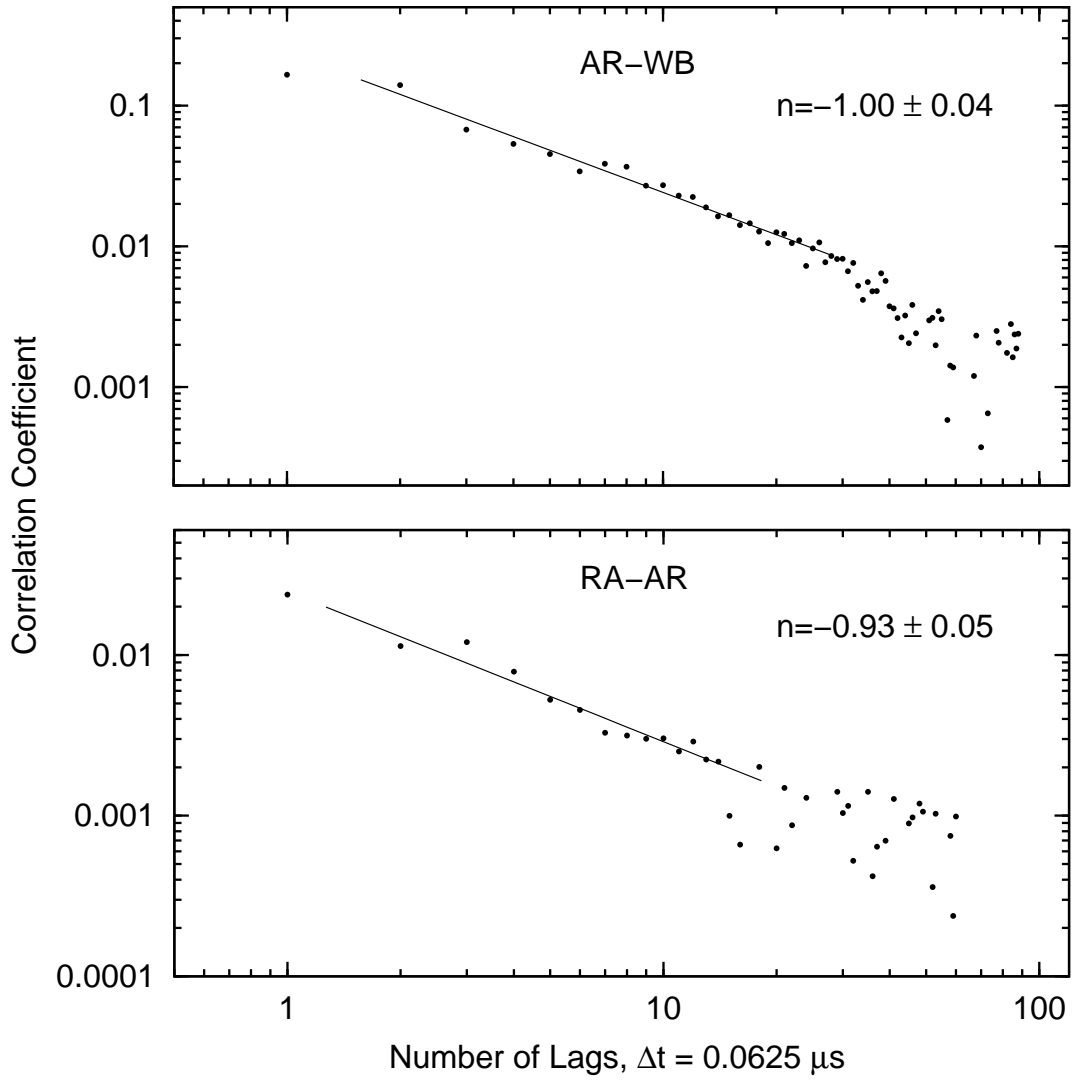} 
 \caption{
Leading part of the coherence function presented in Figure~\ref{fig.9} shown on a log-log scale. The noise 
level has been subtracted. Straight lines correspond to a power-law fit using points not contaminated by noise.
\label{fig.10}}
\end{figure}

We calculated the mean modulus of the coherence function (CF) by averaging the inverse Fourier transforms of the complex 
spectra over the full observation. While the structure function gives us statistically reliable information about the small-scale frequency structure caused by the far layer of scattering plasma, the CF provides detailed information about the nearby layer. The time delay $\tau$ in $\mu$s corresponds to $1/\Delta f$, 
where $\Delta f$ is a frequency shift in spectra expressed in MHz. The limiting resolution in the time delay is determined by the recorded bandwidth and is 0.0625 $\mu$s for our data. The mean CFs for Arecibo-Westerbork (top) and RadioAstron-Arecibo (bottom) are shown in 
Figure~\ref{fig.9}. Note that, for a few individual pulses, the phase of maximal CF differs from the rest of the pulses. This discrepancy may reflect problems with the correlator, so we only averaged the CFs for pulses with the same maximum positions. 

The CF has two components: a narrow, unresolved one corresponding to unscattered emission and a wide, symmetric 
component corresponding to scattered emission, which takes the form $\langle | P_I(\tau)|\rangle\sim\tau^{-(\alpha+1)/2}$ for a power-law phase spectrum (see Equations~(\ref{q34} and \ref{q35})).
Variations over small lags $\tau$ reflect the influence of the nearby layer while variations over larger lags correspond to effects from the far layer. The symmetric structure of scattered component indicates strong angular refraction and demonstrates that the scattering angle is less 
than the refractive angle. 

Figure~\ref{fig.10} shows the temporal CF after normalization by the maximum value. The effects of noise were estimated via a mean of the last 20 points and were then subtracted at quadrature. 
The lines represent the best fitting power laws for points not corrupted by noise; the fit indices were $n_1 = 1.00 \pm 0.04$ for Arecibo-Westerbork and $n_2 = 0.93 \pm 0.05$ for RadioAstron-Arecibo. 
Hence, the power spectra of the electron density fluctuations are similar for the two layers, with an index of $\gamma = 3.00\pm 0.08$ (see 
Equations~(\ref{eq:phase_structure_fns}, \ref{q34} and \ref{q35})). This value is in good agreement with that of 
$\gamma=3.00\pm 0.05$ obtained by \citet{sm2} from analysis of scintillations at 40 to 112 MHz.

\section{Conclusions}


We carried out successful RadioAstron space-ground VLBI
observations of \object{PSR~B0950+08} on January 25, 2012, at
92~cm with a spacecraft distance of 330{,}000 km and projected
interferometer baseline of 220{,}000 km. These measurements
represent the highest angular resolution ever achieved in meter
wavelength observations. The qualitative difference of the form of the structure function
between long and short baselines, as shown in Figure\ \ref{fig.5}
and discussed in Section\ \ref{sec:comparison_with_theory},
suggests the presence of two scattering plasma layers
along the line of sight to the pulsar.
From analysis of the time and frequency scales of the scintillation,
we find that these are located at distances of 4.4 to 16.4\
pc, and of 26 to 170\ pc, as discussed in Section\ \ref{sec:scales_and_distances}.
The nearby layer dominates the temporal structure of the scintillation, while both the nearby and far layers influence the frequency structure of the scintillation.
The velocity of the line of sight in the far and nearby layers, projected onto the direction of
refraction by the ``cosmic prism'' are 20~km\,s$^{-1}$ and $-8$~km\,s$^{-1}$ correspondingly.
After correction for the velocities of the Earth and the pulsar, these are in accord with the velocities typical for interstellar clouds.
The cosmic prism is described below and in Sections\ \ref{sec:cosmicprism}, \ref{sec:shifts_in_freq_and_time}, \ref{sec:scales_and_distances}, and\ \ref{sec:asymmetry}.

The distance to the far screen suggests that it may be located at the outer wall of the Local Bubble,
which lies at about that distance in the direction of \object{PSR~B0950+08} \citep{snow,lal}.
The distance of the nearby screen suggests that it lies at the ionized surface of a nearby molecular cloud;
indeed, such a screen is seen in the direction of the pulsar \citep{Lin}.

From analysis of the temporal coherence function on short and long baselines,
introduced theoretically in Section\ \ref{sec:tempcoherencefunction}
and computed for our observations in Section\ \ref{ref:sec_coherence_function_measurement},
we studied the spectrum of density fluctuations in the two scattering layers.
The spectra of
density fluctuations for the two layers were found to follow power
laws, with indices $\gamma_1=\gamma_2=3.00\pm 0.08$. These
indices differ from the Kolmogorov value of $\gamma = 11/3$.
Note that the Kolmogorov spectrum describes more distant scattering
media very well. However, our results suggest that nearby material has a flatter spectrum.

We observe evidence for refraction by an interstellar plasma wedge, or ``cosmic prism''.
This refraction results in the observed moderate modulation by scintillation of $m=\Delta I/I <1$,
in combination with narrow fractional scintillation bandwidth $\Delta \nu/\nu_0 < 1$.
Usually $m<1$ is characteristic of weak scintillation, whereas $\Delta \nu/\nu_0 < 1$ is characteristic of strong scintillation.
However, if the characteristic value of the refraction angle $\theta_0$ by the cosmic prism is greater than the characteristic value of the diffractive or scattering angle, $\theta_0 >> \Theta_{scat}$, then the frequency structure of the scintillation is formed by
the frequency dependence of the displacement of the beam path, and these two conditions appear together.
We describe this theoretical picture in Sections\ \ref{sec:cosmicprism} and\ \ref{sec:shifts_in_freq_and_time}, and compute parameters of the inferred prism 
from our observations in Sections\ \ref{sec:scales_and_distances} and\ \ref{sec:asymmetry}.
For \object{PSR~B0950+08}, we evaluated the angle of refraction as $\theta_0 = (1.1 - 4.4)$ mas. The refraction is in a direction nearly perpendicular to the velocity of observer.



\acknowledgments

The RadioAstron project is led by the Astro Space Center of the Lebedev Physical Institute of the 
Russian Academy of Sciences and the Lavochkin Scientific and Production Association under a contract 
with the Russian Federal Space Agency, in collaboration with partner organizations in Russia and other countries.
We would like to thank Topasi Ghosh and Chris Salter for their effort to conduct the observation at
Arecibo Radio Telescope. 
The Arecibo Observatory is operated by SRI International under a cooperative agreement 
with the National Science Foundation (AST-1100968) and in alliance with 
Ana G. M\'{e}ndez-Universidad Metropolitana and the Universities Space Research Association.
This research is partly based on observation with the 100-m telescope of the 
MPIfR (Max-Planck-Institut f\"{u}r Radioastronomie) at Effelsberg. 
We are very grateful to the staff at the Westerbork synthesis array for their support.
The data were correlated
at the ASC correlator \citep{kar}.
This work was supported by the Russian Foundation for Basic Research (projects 
12-02-00661 and 13-02-00460) 
and Research Program OFN-17 and  ``The Origin, 
Structure, and Evolution of Objects in the Universe.'' of the Division of Physics, 
Russian Academy of Sciences.
The study was support by the Ministry of Education and Science of Russian Federation,
project 8405.
MJ and CG acknowledge support of the US National Science Foundation (AST-1008865).

{\it Facilities:} \facility{RadioAstron, Arecibo, Effelsberg, WSRT}






\appendix
\section{Appendix}
\label{sec:appendix_a}
For a signal measured in the presence of significant noise, it is advantageous to measure fluctuations in the squared modulus of the cross-power spectrum,
the response of the interferometer. 
We can write the cross-power spectrum as
\begin{eqnarray}
F({\boldsymbol\rho},{\boldsymbol\rho}+\Delta {\boldsymbol\rho},f,t)=V({\boldsymbol\rho},{\boldsymbol\rho}+\Delta {\boldsymbol\rho},f,t)+\nonumber\\
N({\boldsymbol\rho},{\boldsymbol\rho}+\Delta {\boldsymbol\rho},f,t).
\end{eqnarray}
Here, $N$ represents additive white noise from backgrounds. 
Then, to statistically analyze the fluctuations caused by 
interstellar scintillation, we form the function
\begin{eqnarray}
\begin{array}{l}
{\mathcal F}({\boldsymbol\rho},{\boldsymbol\rho}+\Delta {\boldsymbol\rho},f,t) \\
\quad =F({\boldsymbol\rho},{\boldsymbol\rho}+\Delta {\boldsymbol\rho},f,t)F^{\ast}({\boldsymbol\rho},{\boldsymbol\rho}+\Delta {\boldsymbol\rho},f,t+\delta t_1 )\\
\quad =V({\boldsymbol\rho},{\boldsymbol\rho}+\Delta {\boldsymbol\rho},f,t)V^{\ast}({\boldsymbol\rho},{\boldsymbol\rho}+\Delta {\boldsymbol\rho},f,t+\delta t_1 )+\delta_N,
\end{array}
\label{eq::scint_response}
\end{eqnarray}
where $\delta t_1$ is a small shift in time, and $\delta_N$ is given by
\begin{eqnarray}
\begin{array}{l}
\delta_N=N({\boldsymbol\rho},{\boldsymbol\rho}+\Delta {\boldsymbol\rho},f,t)N^{\ast}({\boldsymbol\rho},{\boldsymbol\rho}+\Delta {\boldsymbol\rho},f,t+\delta t_1)\\
\quad +V({\boldsymbol\rho},{\boldsymbol\rho}+\Delta {\boldsymbol\rho},f,t)N^{\ast}({\boldsymbol\rho},{\boldsymbol\rho}+\Delta {\boldsymbol\rho},f,t+\delta t_1)\\
\quad +V^{\ast}({\boldsymbol\rho},{\boldsymbol\rho}+\Delta {\boldsymbol\rho},f,t+\delta t_1)N({\boldsymbol\rho},{\boldsymbol\rho}+\Delta {\boldsymbol\rho},f,t).
\end{array}
\end{eqnarray}
In the case of weak scintillation, Equation\ (\ref{eq::scint_response}) reduces to 
\begin{eqnarray}
\begin{array}{l}
{\mathcal F}({\boldsymbol\rho},{\boldsymbol\rho}+\Delta {\boldsymbol\rho},f,t)\approx  H(t)H(t+\delta t_1) \\
\quad \times [1+j({\boldsymbol\rho},{\boldsymbol\rho},f,t)+j({\boldsymbol\rho}+\Delta {\boldsymbol\rho},{\boldsymbol\rho}+\Delta {\boldsymbol\rho},f,t)]+\delta_N.
\end{array}
\end{eqnarray}
The introduction of $\delta t_1$ decorrelates noise, without significant influence on $j$, or $V$.
The structure function for ${\mathcal F}({\boldsymbol\rho},{\boldsymbol\rho}+\Delta {\boldsymbol\rho},f,t)$ fluctuations can then be written : 
\begin{eqnarray}
D_{\Delta{\mathcal F}}(\Delta {\boldsymbol\rho},f,t)\approx  2H^2D_{\Delta | V|}(\Delta {\boldsymbol\rho},f,t).
\end{eqnarray}




\clearpage

\clearpage

\clearpage

\clearpage


\clearpage

\clearpage

\clearpage

\clearpage

\clearpage

\clearpage
\clearpage

\begin{deluxetable}{lll}
\tablecolumns{3}
\tablewidth{0pc}
\tablecaption{Glossary of Symbols}
\tablehead{
\colhead{Symbol} & \colhead{Description}   & \colhead{Defined}
}
\startdata

$\gamma$                        &  power index of turbulence spectra                                                                          & Sec. \ref{sec:intro} \\

$\nu$                           & observing frequency                                                                                         & Sec. \ref{sub1.1} \\
$\nu_0$                         & center of observing band                                                                                    & Sec. \ref{sub1.1} \\
$f$                             & frequency offset                                                                                            & Sec. \ref{sub1.1}  \\
${\boldsymbol \rho}$            & spatial coordinate in observer plane perpendicular to line of sight                                         & Sec. \ref{sub1.1} \\
$\Delta{\boldsymbol\rho}$       & baseline of interferometer                                                                                  & Sec. \ref{sub1.1} \\

$E$                             & electric field at observer                                                                                  & Eq. \ref{eq:edef} \\
$h(f,t)$                        & electric field of pulsar without propagation, with bandpass                                                 & Eq. \ref{eq:edef} \\
$u({\boldsymbol\rho}f,t)$       & propagation factor                                                                                          & Eq. \ref{eq:edef} \\

$V$                             & interferometric visibility: cross-power spectrum                                                            & Eq. \ref{eq:vdef} \\
$j({\boldsymbol\rho},{\boldsymbol\rho}+\Delta{\boldsymbol\rho},f,t)$                            
                                & propagation factor for $V$                                                                                  & Eq. \ref{eq:jdef} \\
$H(f,t)$                        & flux density of source, with bandpass                                                                       & Eq. \ref{eq:jdef} \\
$\langle .. \rangle_h $         & average over noiselike statistics of source emission                                                        & Sec. \ref{sub1.1} \\

$I$                             & intensity: square modulus of the electric field at a single position                                        & Eq. \ref{eq:Idef} \\
$\Delta I$                      & fluctuations of intensity                                                                                   & Sec. \ref{sec:visbilityandfluctuations} \\

$\ell=1,2$                      & indices for two phase-changing screens                                                                      & Sec. \ref{sec:screensandstatistics} \\
$z_{\ell}$                      & distance of screen $\ell$ from observer                                                                     & Sec. \ref{sec:screensandstatistics} \\
$z$                             & distance of source from observer                                                                            & Sec. \ref{sec:screensandstatistics} \\
${\mathbf x}_{\ell}$              & spatial coordinates in the plane of screen $\ell$                                                           & Sec. \ref{sec:screensandstatistics} \\

$D_{S,\ell}$                    & spatial structure function of phase fluctuations for screen $\ell$                                          & Eq. \ref{eq:Dsdef} \\
$\Phi_{\ell}$                   & screen phase for screen $\ell$                                                                              & Eq. \ref{eq:Dsdef} \\
$\langle .. \rangle_S$          & average over realizations of scattering medium                                                              & Eq. \ref{eq:Dsdef} \\

$\Theta_{scat,\ell}$            & characteristic deflection angle for screen $\ell$                                                           & Eq. \ref{eq:phase_structure_fns} \\
$k$                             & wavenumber                                                                                                  & Eq. \ref{eq:phase_structure_fns} \\
$\lambda$                       & wavelength                                                                                                  & Eq. \ref{eq:phase_structure_fns} \\
$\alpha_{\ell}$                 & power-law index of structure function for screen $\ell$                                                     & Eq. \ref{eq:phase_structure_fns} \\
$m$                             & modulation index of scintillation                                                                           & Sec. \ref{sec:screensandstatistics} \\

${\boldsymbol\theta}_0$         & refraction angle of cosmic prism                                                                            & Sec. \ref{sec:cosmicprism} \\

${\boldsymbol\theta}_{f}$       & angular displacement of source by cosmic prism at frequency $f$                                             & Eq. \ref{q15} \\
${\boldsymbol\rho}_{f,\ell}$    & displacement in observer plane of scintillation pattern of screen $\ell$,                                   & Eqs. \ref{q16},\ref{eq:f_shift_2} \\
                                &  caused by cosmic prism                                                                                     &  \\
${\boldsymbol\rho}_{t,\ell}$    & spatial displacement of observer relative to                                                                & Eqs. \ref{eq:time_shift_1},\ref{eq:time_shift_2} \\
& scintillation pattern of screen $\ell$ & \\
${\mathbf V}_{obs}$               & observer's velocity transverse to the line of sight                                                         & Eqs. \ref{eq:time_shift_1},\ref{eq:time_shift_2} \\
${\mathbf V}_{scr,\ell}$          & velocity of screen $\ell$ transverse to the line of sight                                                   & Eqs. \ref{eq:time_shift_1},\ref{eq:time_shift_2} \\
${\mathbf V}_{PSR}$               & velocity of pulsar transverse to the line of sight                                                          & Eq. \ref{eq:time_shift_2} \\
${\mathbf V}_{\ell}$              & observer's resultant velocity relative to scintillation pattern of screen $\ell$                            & Eqs. \ref{eq:time_shift_1},\ref{eq:time_shift_2} \\

$D_{\Delta I}$                  & structure function of intensity variations in observer plane                                                & Eq. \ref{eq:defineIstructurefunc} \\
$D_{\Delta I,\ell}$             & structure function of intensity variations in observer plane, for screen $\ell$                             & Eq. \ref{eq:screenIstructurefunc} \\
$D_{\Delta |j|,\ell}$           & structure function of intensity variations from scintillation, for screen $\ell$                            & Eq. \ref{q18} \\
${\mathbf r}_{\ell}$              & generalized position variable including effects of refraction and                                           & Eq. \ref{eq:r_def} \\
                                & motion of the scintllation pattern, for screen $\ell$                                                       &  \\
$\beta_l$                       & angle between direction of dispersion of cosmic prism ${\boldsymbol\theta}_0$ and                           & Eq. \ref{sec:time_freq_baseline} \\
                                & velocity  ${\boldsymbol V}_{\ell}$ of observer relative to scintillation pattern of screen ${\ell}$         &  \\
$\varphi$                       & angle between ${\boldsymbol\theta_0}$ and baseline $\Delta{\boldsymbol\rho}$                                & Eq. \ref{sec:time_freq_baseline} \\

$\rho_{Fr,\ell}$                & Fresnel spatial scale for screen $\ell$                                                                     & Eqs. \ref{eq:Fresnel_close},\ref{eq:Fresnel_far} \\
$f_{Fr,\ell}$                   & Fresnel frequency scale for screen $\ell$                                                                   & Eqs. \ref{eq:fFr1_def},\ref{eq:fFr2_def} \\
$t_{Fr,\ell}$                   & Fresnel time scale for screen $\ell$                                                                        & Eqs. \ref{eq:tFr1_def},\ref{eq:tFr2_def} \\

$P$                             & temporal coherence function                                                                                 & Eq. \ref{eq:tempcoherfuncdef} \\
$\tau$                          & time lag of coherence function                                                                              & Eq. \ref{eq:tempcoherfuncdef} \\
$P_H$                           & temporal coherence function of source                                                                       & Eq. \ref{eq:PHdef} \\
$B_u$                           & spatial coherence function of scattering                                                                    & Eq. \ref{q29} \\
$P_0$                           & temporal coherence function of unscattered emission                                                         & Sec. \ref{sec:tempcoherencefunction} \\
$P_{S,\ell}$                    & contribution of screen $\ell$ to temporal coherence function                                                & Eq. \ref{eq:PS1def},\ref{q34},\ref{q35} \\

$\tau_{Fr}$                     & time lag of coherence function corresponding to $1/f_{Fr}$                                                  & Eq. \ref{q34},\ref{q35} \\
$H_0$                           & flux density of source integrated over frequency                                                            & Eq. \ref{eq:PS1def} \\

$q_{\perp}, q_{||}$             & components of spatial frequency in plane of screen,                                                         & Eq. \ref{eq:PS1def} \\
                                & parallel and perpedicular to refraction angle ${\boldsymbol \theta}_0$                                      & \\

$\langle ... \rangle_{f,t}$ & average over time and frequency & Eq.\ \ref{eq:CCF_def} \\
${\rm CCF} (\Delta t)$ & cross-correlation at time lag $\Delta t$ & Eq.\ \ref{eq:CCF_def} \\

$t_{SC}$ & timescale of scintillation & Sec.\ \ref{sub4.1} \\

$F$ & normalized spectrum & Eq.\ \ref{eq:Fdef} \\
${\mathcal F}$ & structure function & Eq.\ \ref{eq:structfuncdef} \\
$\Delta f_0$ & maximum frequency difference sampled & Eq.\ \ref{q42} \\

${\mathcal D}(\Delta f, \Delta t)$ & quantified ratio of asymmetry of structure functions & Eq.\ \ref{eq:asymmetry_ratio} \\

\enddata
\end{deluxetable}


\end{document}